\documentclass[11pt]{article} 
	\usepackage[margin=1.25 in, bottom=1.3 in]{geometry} 
	\usepackage[english]{babel} 
	\usepackage{authblk}
	\usepackage{amsmath, amssymb, textcomp, amsthm, mathtools,mathrsfs}  
	\usepackage{algorithm}
	\usepackage{algpseudocode}
	\usepackage{esint}
	\usepackage{tensor}
	\usepackage{blkarray}
	\usepackage{wasysym}
	\usepackage{stmaryrd}
	\usepackage{dsfont}
	\usepackage{hyperref}
   	\usepackage{natbib}
	\usepackage[utf8]{inputenc} 
	\usepackage{csquotes}
	\usepackage{footnote,footmisc} 
	\usepackage{setspace} 
	\usepackage{threeparttable}
	\usepackage{multirow}
	\usepackage{tikz}
	\usetikzlibrary{calc}

	\tikzset{cross/.style={cross out, draw=black, minimum size=2*(#1-\pgflinewidth), inner sep=0pt, outer sep=0pt},
cross/.default={2pt}}
	\usepackage{pgfplots}
	\usetikzlibrary{shapes.misc}
	\usetikzlibrary{patterns}
	\usepackage{xcolor}
	\hypersetup{
    	colorlinks,
    	linkcolor={red!50!black},
    	citecolor={blue!50!black},
    	urlcolor={blue!80!black}
	}
	\usetikzlibrary{decorations.pathreplacing}
	\usetikzlibrary{arrows}
	\usepackage{float} 
	\newcommand{\GG}[1]{}
   	 \allowdisplaybreaks
	\usepackage[toc,page]{appendix}

	\newtheorem{theorem}{Theorem}
	 
	\newtheorem{proposition}{Proposition} 
	
	\newtheorem*{conjecture*}{Conjecture}
	\newtheorem{definition}{Definition}

	\newtheorem{assumption}{Assumption}
	\usepackage{graphicx}  
	\usepackage{marvosym} 
	\makeatletter
\def\moverlay{\mathpalette\mov@rlay}
\def\mov@rlay#1#2{\leavevmode\vtop{%
   \baselineskip\z@skip \lineskiplimit-\maxdimen
   \ialign{\hfil$\m@th#1##$\hfil\cr#2\crcr}}}
\newcommand{\charfusion}[3][\mathord]{
    #1{\ifx#1\mathop\vphantom{#2}\fi
        \mathpalette\mov@rlay{#2\cr#3}
      }
    \ifx#1\mathop\expandafter\displaylimits\fi}
\makeatother

    \DeclareMathOperator*{\argmin}{\arg\!\min}

    \newtheorem*{assumptions*}{\assumptionnumber}
\providecommand{\assumptionnumber}{}


\title{Distributional synthetic controls}
\author{Florian F Gunsilius}
\date{\today}
\affil{University of Michigan}
\begin{document}
\maketitle
\begin{abstract}
This article extends the widely-used synthetic controls estimator for evaluating causal effects of policy changes to quantile functions. The proposed method provides a geometrically faithful estimate of the entire counterfactual quantile function of the treated unit. Its appeal stems from an efficient implementation via a constrained quantile-on-quantile regression. This constitutes a novel concept of independent interest. The method provides a unique counterfactual quantile function in any scenario: for continuous, discrete or mixed distributions. It operates in both repeated cross-sections and panel data with as little as a single pre-treatment period. The article also provides abstract identification results by showing that any synthetic controls method, classical or our generalization, provides the correct counterfactual for causal models that preserve distances between the outcome distributions. Working with whole quantile functions instead of aggregate values allows for tests of equality and stochastic dominance of the counterfactual- and the observed distribution. It can provide causal inference on standard outcomes like average- or quantile treatment effects, but also more general concepts such as counterfactual Lorenz curves or interquartile ranges. 
\\

\end{abstract}

\section{Introduction}
The method of synthetic controls, introduced in \citet*{abadie2003economic} and \citet*{abadie2010synthetic}, has become a main tool for estimating causal effects in comparative case studies with aggregate interventions and a limited number of large units. It is designed for settings where some units are subject to a policy intervention and others are not; the respective outcomes of interest are measured in each population before and after the policy intervention, potentially for many periods. The control groups are used to account for unobserved trends in the outcome over time that are unrelated to the effect of the policy intervention. The insight is that an optimally weighted average of the available potential control groups, the \emph{synthetic control group}, often provides a more appropriate comparison than a single control group alone \citep*{abadie2019using}. The method has been applied in many settings, such as the analysis of the decriminalization of indoor prostitution \citep*{cunningham2017decriminalizing}, the minimum wage debate (\citeauthor*{allegretto2017credible} \citeyear{allegretto2017credible}, \citeauthor*{jardim2017minimum} \citeyear{jardim2017minimum}, \citeauthor*{neumark2017reply} \citeyear{neumark2017reply}), and immigration (\citeauthor*{borjas2017wage} \citeyear{borjas2017wage}, \citeauthor*{peri2019labor} \citeyear{peri2019labor}); see \citet*{abadie2019using} for further references. 

The original method of synthetic controls is designed for settings with aggregate scalar- or vector-valued quantities, in which linear regression approaches are not applicable because of data limitations \citep*{abadie2019using}. Researchers and policy makers are frequently interested in estimating the causal impacts of policy interventions for different quantiles of the population, however. An example is assessing the effects of minimum wage policies on the overall income distribution. An excellent illustration for the relevance of estimating distribution functions is \citet*{ropponen2011reconciling} who applies the changes-in-changes estimator by \citet*{athey2006identification} to estimate the effect of minimum wage changes on employment levels in order to reconcile the classical conflicting results about the minimum wage debate in \citet*{card2000minimum} and \citet*{neumark2000minimum}. 

In this article we introduce the distributional synthetic controls method, an extension of the classical method of synthetic controls from aggregate values to quantile functions on the real line. In this case the classical synthetic controls estimator reduces to a constrained linear regression (\citeauthor*{abadie2015comparative} \citeyear{abadie2015comparative}, \citeauthor*{doudchenko2016balancing} \citeyear{doudchenko2016balancing}) and our method provides a natural extension of this method. The idea of constructing a weighted average of control units that replicates the geometric properties of the target distribution as closely as possible is especially worthwhile in these settings. For instance, in the classical example of a policy change at the state-level \citep*{abadie2010synthetic} our method allows to obtain the entire counterfactual distribution of the treated state after the policy change instead of merely aggregate values. What is more, it allows for causal inference on quantiles of the distribution or any function of the counterfactual distribution such as Lorenz curves \citep*{gastwirth1971general}. This could be important for the research on inequality.

\begin{figure}[h!]
\centering
\textbf{Example of geometric faithfulness of Wasserstein barycenters vs mixtures }
\includegraphics[width=7.5cm,height=6cm]{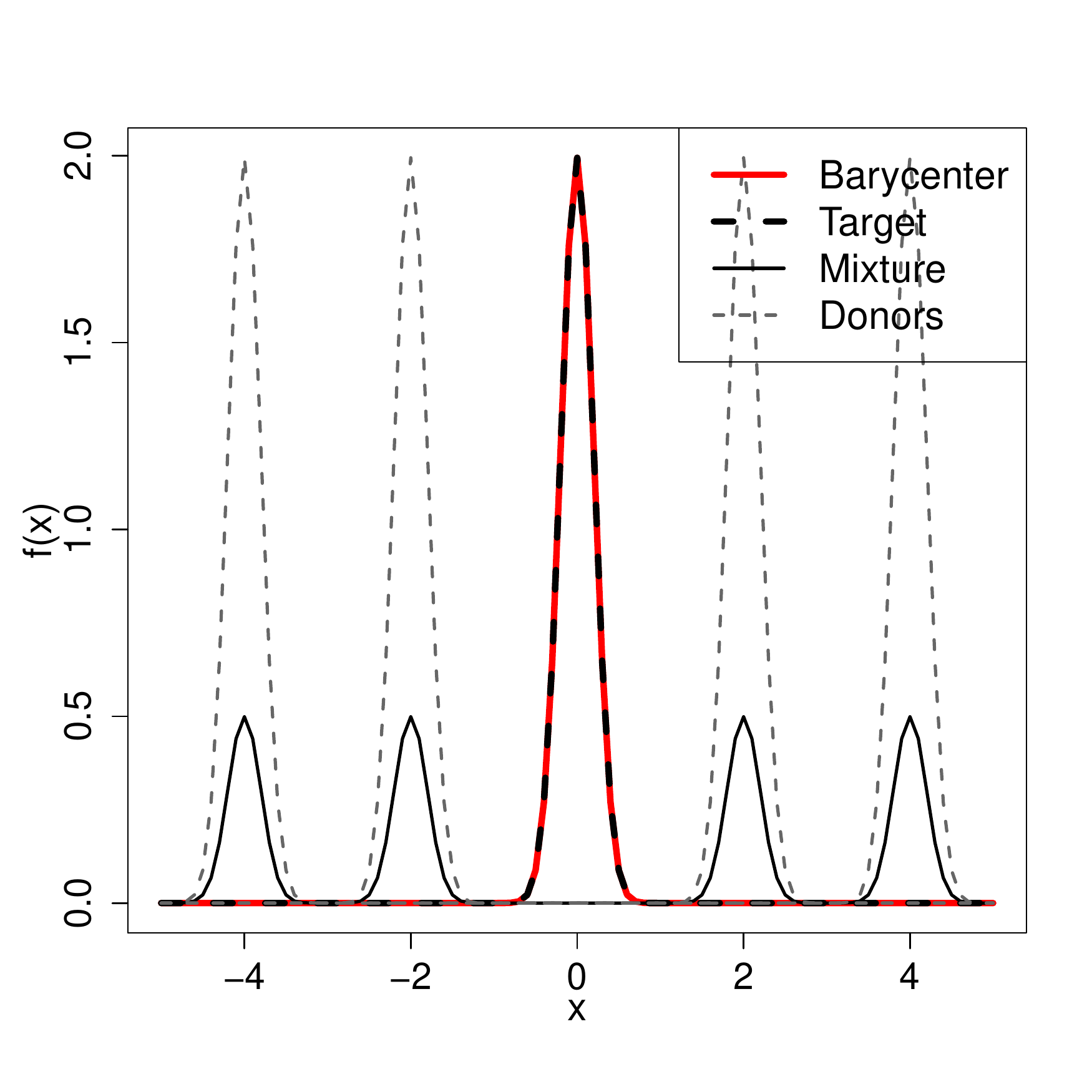}
\includegraphics[width=7.5cm,height=6cm]{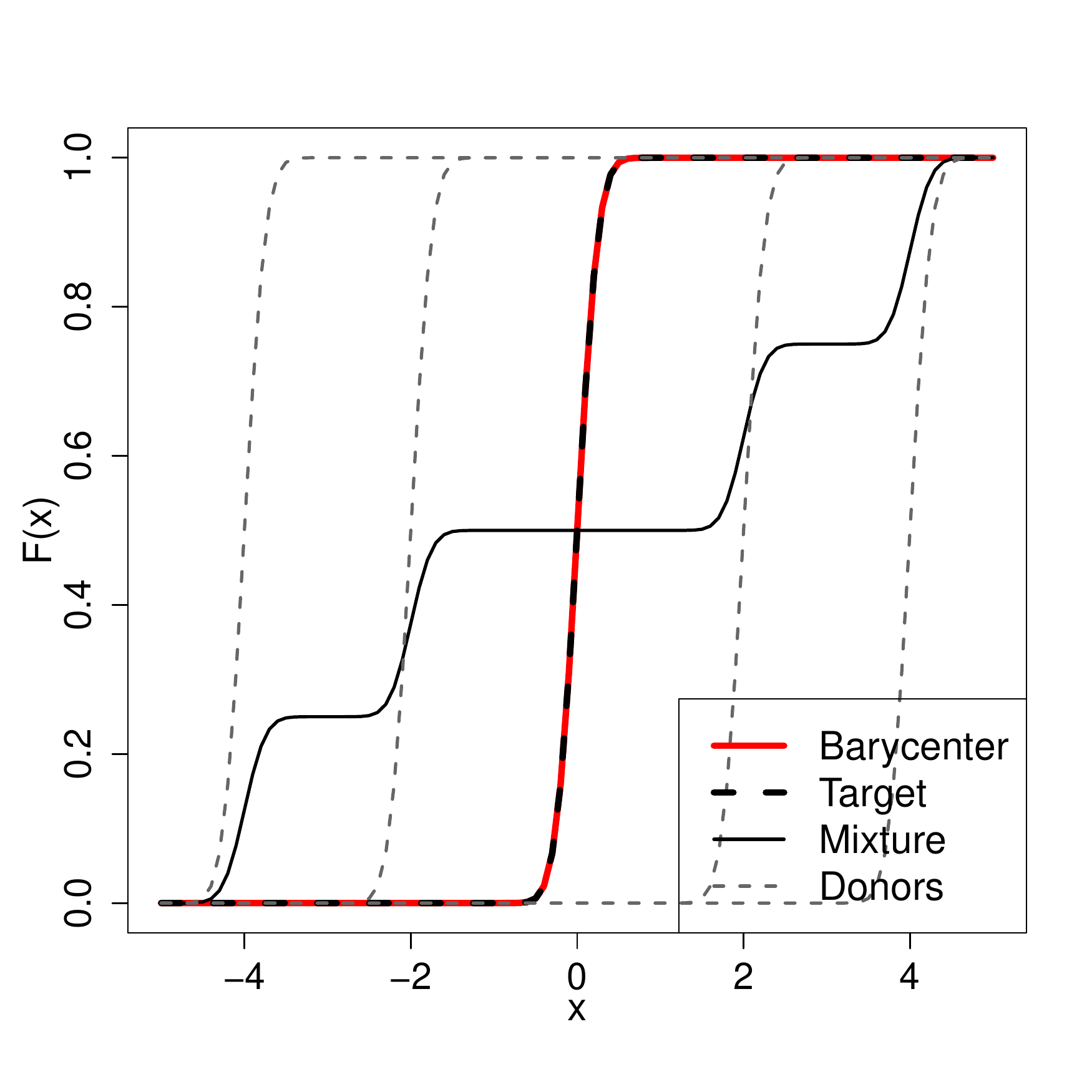}
\caption{Four control distributions $F_j(x)$ are univariate Gaussians (gray dashed) with common variance $\sigma^2=0.04$ and means $\mu_1 = -4$, $\mu_2 = -2$, $\mu_3 = 2$, $\mu_4 = 4$. Target distribution to replicate (black dashed) is Gaussian with variance $\sigma^2=0.04$ and mean $\mu=0$. Wasserstein barycenter with weights $\tfrac{1}{4}$ for each donor distribution (red) replicates the target perfectly. A linear mixture $\frac{1}{4}\sum_{j=1}^4 F_j$ is multimodal and does not replicate the target. Left panel: densities, right panel: cumulative distribution functions.}\label{normalplot}
\end{figure}

The key concept for our method is a generalization of the linear average of the aggregate outcomes from the classical synthetic control method to appropriate concepts of averages between distribution functions. We argue that the appropriate generalization is to barycenters\footnote{Barycenters, or Fr\'echet means \citep*[chapter 9]{kendall2009shape}, are the natural generalization of linear averages to general metric spaces. Below we introduce these notions formally.} defined on the space of distribution functions with finite second moments, equipped with the $2$-Wasserstein metric\footnote{In different areas the Wasserstein distance is known under the name Mallows distance or earth mover's distance \citep*{levina2001earth}.} \citep*[chapter 6]{villani2008optimal}, or ``Wasserstein barycenters'' for short \citep*{agueh2011barycenters}. The reason for choosing the Wasserstein barycenter instead of classical linear mixtures is that we want a method that is ``geometrically faithful'' in replicating the target distribution. This means that it attempts to replicate all geometric properties of the target distribution, such as the support, the moments, the quantiles, and so on. In contrast, standard linear mixture models do not replicate the support or quantiles of the target and are therefore not geometrically faithful as depicted in Figure \ref{normalplot}. 

The goal of a synthetic controls method is to obtain a synthetic unit that replicates all quantiles of the target as closely as possible. Wasserstein barycenters provide the framework for this. An example is depicted in Figure \ref{normalplot}, where one wants to find an optimal combination of $4$ Gaussian distributions $\left\{F_{j}\right\}_{j=2,\ldots,5}$ (gray dashed) with the same small variance and different means that replicate as closely as possible the target distribution $F_1$ (black dashed) which is another Gaussian with the same variance but a different mean. The Wasserstein barycenter with equal weights for each donor distribution is geometrically faithful in that it replicates the target distribution perfectly. In contrast, a classical linear mixture approach $\bar{F} = \frac{1}{4}\sum_{j=2}^5 F_j$ is not geometrically faithful: it is a multimodal distribution that does not replicate any part of the target distribution. This is true in any setting, not just for Gaussians: Wasserstein barycenters replicate all moments and quantiles of the target (see e.g.~Figure 1 in \citet*{cuturi2014fast} which shows this in a bivariate setting), while standard linear mixtures do not have this property.

The proposed method constitutes a novel concept and is the most natural extension of a linear regression of outcomes to probability measures in Wasserstein space. Note that this differs from the well-known quantile regression \citep*{koenker1978regression}. In classical quantile regression one regresses a specific quantile on covariates. Our method regresses quantile functions on whole quantile functions. The difference is that the explanatory variables in our case are also functional quantities, while the explanatory variables in classical quantile regression are finite-dimensional. In this respect, note that \citet*{bonneel2016wasserstein} introduce a similar version to ours, but for color transport on images, i.e.~$2$-dimensional histograms. In this case the simple approach becomes a rather complicated to solve bi-level optimization problem which has many local optima. The authors provide a fixed points approach which only finds a stationary point; a local solution is enough for color transport as the weights are not important. In contrast to color transport, obtaining consistent weights is imperative for a causal inference approach as they are the basis for computing the counterfactual distribution. The proposed method is efficient to implement, works for all type of distributions, and always obtains a unique set of optimal weights. 

The proposed method based on Wasserstein barycenters provides a synthetic control unit at the level of interest: the state-level, not the individual level. That is, it matches distributions globally, i.e.~it finds an optimal weight for every distribution of the control units that replicates the target distribution as closely as possible. In contrast, linear point-wise approaches like \citet*{chen2020distributional} which is based on classical quantile regression or more na\"ive approaches that (i) decompose distributions into bins and (ii) match these bins between the different distributions are local: they obtain different weights for each point in the distribution and hence obtain a ``weight-function''. As a result, these ad-hoc synthetic controls methods are unstable with the choice of points or bins; in particular, they need to make strong assumptions that the optimal weights for each point or bin are ``coherent'' in the sense that they are the same or at least similar within a given state (e.g.~Assumption 1 (ii) in \citet*{chen2020distributional}), something that is rarely satisfied in practice. Such local approaches also cannot provide uniform inference results on the whole counterfactual distribution as they obtain weights that change with each quantile. Finally, these methods implicitly need to assume that each quantile value evolves linearly over time an assumption that is restrictive and rarely met by common stochastic processes. For instance, the transition densities of stochastic differential equations with a nonlinear drift term \citep*{ait1999transition} violate this assumption.

Our method solves all of these issues. First, it provides synthetic control units at the level of interest. Second, it is always geometrically faithful by construction. Third, its implementation is straightforward because it takes the form of a novel linear regression on quantile functions. Fourth, it replicates the whole counterfactual quantile function of the treatment group with a unique set of weights in any setting: for continuous, discrete, or mixed distributions. Fifth, it allows us to provide tests for the counterfactual distribution. 

Furthermore, the method is the natural complementary approach to the changes-in-changes estimator \citep*{athey2006identification} in the same sense as the synthetic controls estimator is a complementary approach to the linear difference-in-differences estimator. In particular, we show that the changes-in-changes estimator relies on the exact same mathematical tools as our synthetic controls estimator. The distributional synthetic controls method has three main advantages over the changes-in-changes estimator: (i) it always provides a unique set of weights and counterfactual quantile function, even if the distributions are not continuous, (ii) it does not require the rank-invariance assumption\footnote{Rank-invariance means that the quantiles between the unobservable and the outcome are preserved. For instance, in the classroom setting where the unobservable is ability of the student, rank invariance implies that the student with the highest unobserved ability will receive the highest outcome, the student with the second highest ability will receive the second highest outcome, and so on. The assumption that one can rank individuals based on a one-dimensional criterion and that this ranking stays fixed between distributions can be strong in some settings, see for instance \citet*[p.~131]{dette2016testing}.} on the causal model \citep*{matzkin2003nonparametric} to produce the correct counterfactual distribution, and (iii) it allows for tests of the entire counterfactual quantile function. 

The fact that our method is geometrically faithful allows us to perform causal inference on the whole counterfactual quantile function or any function based on it. Rejecting the null-hypothesis of equality of quantile functions is a necessary condition for a causal effect. We also provide large sample results for tests of stochastic dominance and a placebo permutation test in analogy to the classical method \citep*{abadie2010synthetic}: apply the method not just to the actual treatment unit, but also each control unit. If there is a true causal effect, then the change between pre-intervention- and post-intervention period should be largest for the treated unit compared to the control units. The novelty is that one can obtain causal inference on any function of the counterfactual distribution such as averages, quantiles, but also more general concepts such as Lorenz curves \citep*{gastwirth1971general}, Gini-coefficients \citep*{gastwirth1972estimation}, interquartile ranges, or other ROC-curves (\citeauthor*{hsieh1996nonparametric} \citeyear{hsieh1996nonparametric}, \citeauthor*{schechtman2019relationship} \citeyear{schechtman2019relationship}).

Another contribution of this article is the analysis of a general set of nonlinear causal models for which our method identifies the correct counterfactual distribution. We analyze the nonlinear causal model introduced for the changes-in-changes estimator \citep*[equation (3)]{athey2006identification} through the lens of our synthetic controls approach. The takeaway for applied researchers is that the synthetic controls method, no matter if classical or our distributional extension, works with a causal model that preserves the distance between the outcomes. This follows from the fact that the synthetic controls method depends on an extrapolation step over time: in the pre-period, one finds an optimal combination of the control groups which matches the treatment group as closely as possible. The identifying assumption for causal inference then is that this combination stays optimal in the post-treatment period.

This article complements recently introduced dynamic and nonlinear implementations of the synthetic controls idea, such as \citet*{abadie2017penalized}, \citet*{amjad2018robust}, \citet*{arkhangelsky2019synthetic}, \citet*{athey2018matrix}, \citet*{athey2018design}, \citet*{ben2018new}, \citet*{chernozhukov2018inference}, \citet*{doudchenko2016balancing}, \citet*{imai2019should}, and \citet*{viviano2019synthetic}. The main difference to these approaches is that our method applies the idea of constructing a synthetic control group directly to probability measures and does not consider an aggregate quantity or functionals of the distribution. Also, our method is concerned with estimating the counterfactual distribution for one well-defined treatment intervention and does not deal with the problem of staggered adoption, which some of the other approaches cover. A natural next step for our method is to extend it to staggered adoption settings. 


\section{A generalization of the synthetic controls model}\label{setupsec}
\subsection{The classical model and approach}
The setup for the proposed method is analogous to the classical synthetic controls approach (\citeauthor*{abadie2003economic} \citeyear{abadie2003economic}, \citeauthor*{abadie2010synthetic} \citeyear{abadie2010synthetic}, \citeauthor*{abadie2019using} \citeyear{abadie2019using}), and the notation is chosen to reflect this fact. Suppose we have data on a set of $J+1$ units, where the first unit $j=1$ is the treated unit and $j=2,\ldots,J+1$ are the potential control units. These units are observed over $T$ time periods, where $T_0<T$ denotes the last time period prior to the treatment intervention in unit $j=1$. In practice, interventions can have an impact prior to their formal implementation, in which case one should define $T_0$ as the last time point before the intervention can have an impact \citep*[p.~494]{abadie2010synthetic}. In accordance with the literature we call $t\leq T_0$ the pre-intervention- or pre-treatment periods and $t> T_0$ the post-intervention- or post-treatment periods.

The classical method of synthetic controls focuses on an aggregated outcome $Y_{jt}$ that is observed for each unit $j=1,\ldots, J$ over the time periods $t=1,\ldots, T$. We denote by $Y_{jt,N}$ the outcome of group $j$ that would be observed at time $t$ in the absence of the intervention; analogously, we denote by $Y_{jt,I}$ the outcome of group $j$ that would be observed at time $t$ if the unit was exposed to the treatment at time $t> T_0$. The standard assumption in this setting is that the intervention has no effect on the outcome before the implementation period, so that we have $Y_{jt,N}=Y_{jt,I}$ for all units $j$ and all pre-intervention periods $t<T_0$. 

The key quantity to estimate is $Y_{1t,N}$, the outcome of the treatment unit had it not received the treatment in the post-intervention periods. Based on this, one defines the effect $\alpha_{jt} = Y_{jt,I}-Y_{jt,N}$ of the intervention for unit $j$ at time $t$, so that one can write the observable outcome in terms of the counterfactual notation as
\[Y_{jt} = Y_{jt,N} + \alpha_{jt} D_{jt},\] where 
\[D_{jt} = \begin{cases} 1&\text{if $j=1$ and $t>T_0$}\\ 0 & \text{otherwise.}\end{cases}\]

The goal in the classical setting is to estimate the treatment effect on the treated group in the post-treatment period, i.e.
\[\alpha_{1t} = Y_{1t,I} - Y_{1t,N} = Y_{1t} - Y_{1t,N}\qquad\text{for $t>T_0$,}\]
which requires a model for the unobservable $Y_{1t,N}$. \citet*{abadie2010synthetic} introduce a linear factor model
\[Y_{jt,N} = \delta_t + \theta_tX_{j}+ w_t\mu_j+\varepsilon_{jt},\] where $\delta_t$ is a univariate factor, $w_t$ is a vector containing factors whose loadings are captured in $\mu_j$, and where $X_j$ are observed covariates of the respective units. The error terms $\varepsilon_{jt}$ are zero mean transitory shocks. The idea for the synthetic controls method is that $Y_{jt,N} = Y_{jt}$ for $t>T_0$ and $j=2, \ldots, J+1$, so that the treatment effect on the treatment group can be obtained by a weighted average
\[\hat{\alpha}_{1t} = Y_{1t} - \sum_{j=2}^{J+1}\lambda^*_jY_{jt},\] where $\{\lambda^*_{j}\}_{j=2,\ldots, J}$ is an optimal set of weights. 

The classical synthetic controls estimator in this setting then proceeds in two stages. 
\begin{enumerate}
\item In the first stage, one obtains the optimal weights $\lambda^*\coloneqq \{\lambda^*_j\}_{j=2,\ldots,J+1}$ which lie in the $(J-1)$-dimensional probability simplex\footnote{The $(J-1)$-dimensional probability simplex $\Delta^{J-1}$ is defined as consisting of $J$-dimensional vectors $\lambda$ whose entries are all non-negative and sum to $1$.} $\Delta^{J-1}$ and are chosen such that they minimize a weighted Euclidean distance 
\begin{equation*}
\|x_1-X_0\lambda\|\coloneqq \left(\sum_{k=1}^K v_k\left(X_{k1} - \lambda_{2}X_{k2}-\ldots - \lambda_{J+1} X_{kJ+1}\right)^2\right)^{1/2},
\end{equation*} 
where $X_0$ is the $k\times J$- matrix of potential covariates corresponding to the $J$ control groups and $v\coloneqq \Delta^{K-1}$ is another set of weights which needs to be chosen by the researcher. \citet*{abadie2003economic}, \citet*{abadie2010synthetic}, and \citet*{abadie2015comparative} provide possible choices for $v$. 
\item In the second stage, the obtained optimal weights $\{\lambda^*_j\}_{j=2,\ldots,J+1}$ from this minimization are used to create $\hat{Y}_{1t}^N$ in the post-treatment periods as 
\begin{equation}\label{classicaleq}
\hat{Y}_{1t}^N = \sum_{j=2}^{J+1} \lambda_j^* Y_{jt},\qquad\text{for $t>T_0$,} 
\end{equation} 
based on which one can estimate $\hat{\alpha}_{1t} = Y_{1t} - \hat{Y}_{1t}^N$. 
\end{enumerate}

\subsection{The distributional setting}
The idea for the distributional setting considered in this article is analogous to the classical setting, but with $F_{Y_{jt}}$, the probability law of $Y_{jt}$, as the quantity of interest. The idea is identical: we want to approximate the counterfactual distribution $F_{Y_{1t},N}$ by an optimally weighted average of the control distributions $\{F_{Y_{jt}}\}_{j=2,\ldots,J+1}$ that replicates the geometric properties of the target distribution as closely as possible. We choose the Wasserstein barycenter for this task. In particular, we have already argued in the introduction that the classical linear mixture distribution is not a good candidate, as it is not ``geometrically faithful''. 

Our idea is to instead focus on the associated quantile functions instead of the distributions and to take linear averages of those:
\[\sum_{j=2}^{J+1}\lambda_j F^{-1}_{Y_{jt}}(q)\qquad\text{for all $q\in[0,1]$},\]
where the quantile function is defined in the usual way as
\[F^{-1}(q)\coloneqq\inf_{x\in\mathbb{R}}\left\{F(x)\geq q\right\}.\]
This corresponds to the $2$-Wasserstein barycenter in the space of quantile functions and provides the natural extension of the classical synthetic controls estimator. We show this formally in section \ref{methodsection}. Note that the weights $\lambda_j$ do not depend on the quantile, which is the main difference to existing approaches like \citet*{chen2020distributional} or more na\"ive point-wise approaches: the proposed method provides a synthetic control unit at the level of interest, the state-level, not the individual level. This comes with all the advantages mentioned in the introduction.

We now introduce the causal model for the outcome of the groups without treatment, analogous to the linear factor model in the classical setting. This model will be the same causal model as used in the changes-in-changes estimator \citep*{athey2006identification}. In the setting of a few repeated cross-sections, we assume that unobserved heterogeneity $U_j$ for each unit $j$ does not change over time, which leads to the nonseparable model \citep*[equation (3)]{athey2006identification}
\begin{equation}\label{modeleq}
Y_{jt,N} = h(t,U_{j}), \qquad \forall t, \quad j=1,\ldots,J+1.
\end{equation}
Model \eqref{modeleq} is nonseparable in the unobservable $U_j$ and the time period. Here, $U_{j}$ denotes the unobservable characteristics of unit $j$. $h$ is some measure-preserving function which is often called a ``production function''. Equation \eqref{modeleq} captures the idea that the outcome of a unit $j$ with $U_j=u$ will be the same in a given time period. 
The number of time periods can be as low as $2$, with one pre-treatment period and one post-treatment period. In contrast, the classical factor model of the synthetic controls estimator requires an infinite number of pre-treatment periods in order to identify the factors in their model. 
Model \eqref{modeleq} encompasses the classical linear difference-in-differences model by setting $U_{ij} = \alpha +\gamma\cdot\mathds{1}\{j=1\} +\varepsilon_i$, where $\mathds{1}\{\cdot\}$ is the indicator function taking the value $1$ if the statement in braces is true and $0$ otherwise.

Our setting can also allow for a more general model in which the unobservables are allowed to change over time:
\begin{equation}\label{modeleq1}
Y_{jt,N} = h(t,U_{jt}), \qquad \forall t, \quad j=1,\ldots,J+1.
\end{equation}
Model \eqref{modeleq} is probably appropriate in settings with short time horizons, i.e.~in repeated cross-sections: in this case it is unlikely that the unobservables will change drastically. Model \eqref{modeleq1} is more appropriate in a panel setting and for longer time horizons. Our method, just like the changes-in-changes estimator, works in both settings. 

The whole question for identification of the counterfactual distribution in this model is what type of functions $h$ provide the correct counterfactual distribution when applying our estimator. In section \ref{identsection} we show that one obtains the correct counterfactual for functions $h$ that preserve the Wasserstein distances between the distribution of the unobservables and the outcomes. This is a novel identification result for synthetic control estimators, as we derive it from a nonseparable and nonlinear model. 

\section{The method of distributional synthetic controls}\label{methodsection}
\subsection{The $2$-Wasserstein space and the $2$-Wasserstein barycenter on the real line}\label{toolssection}
Before presenting our method, we briefly need to introduce the necessary mathematical machinery, in particular the $2$-Wasserstein space. The $2$-Wasserstein space is a metric space on the set of all probability distributions defined on a general space such as the Euclidean space. In this article we only consider distributions on the real line. The $2$-Wasserstein distance $W_2^2(P_1,P_2)$ between two probability measures $P_1$ and $P_2$ with corresponding distributions $F_1, F_2$ with finite second moments is defined as \citep*[Theorem 2.18]{villani2003topics}:
\[W_2(P_1,P_2) = \left(\int_0^1\left\lvert F^{-1}_1(q) - F^{-1}_2(q)\right\rvert^2 dq\right)^{1/2},\] where $F_1^{-1}$ is the quantile function corresponding to $P_1$ and $F_2^{-1}$ is the quantile function corresponding to $P_2$. It is a well-known fact that the quantile function of the $2$-Wasserstein barycenter is simply the weighted average of the quantile functions of the donor distributions (\citeauthor*{agueh2011barycenters} \citeyear{agueh2011barycenters}, \citeauthor*{bigot2017geodesic} \citeyear{bigot2017geodesic}), i.e.~for the barycenter $\bar{P}$ of distributions $\{P_{Y_{jt}}\}_{j=2,\ldots,J+1}$ with weights $\lambda = (\lambda_2,\ldots,\lambda_{J+1})$
\begin{equation}\label{quantilebar}
F_{\bar{P}}^{-1}(q) = \sum_{j=2}^{J+1}\lambda_j F_{Y_{jt}}^{-1}(q)\qquad\text{for all $q\in[0,1]$}.
\end{equation}
The $2$-Wasserstein space in the univariate setting is hence a linear space on quantile functions. 

\subsection{Distributional synthetic controls}
Based on the above concepts, we now introduce the method of synthetic controls. It proceeds exactly as in the classical setting, except that we now assume that we observe the distribution functions $F_{Y_{jt}}$ for $j=1,\ldots, J+1$ instead of aggregated outcomes. 

The method of distributional synthetic controls then proceeds in perfect analogy to the classical method. At every time point $t\leq T_0$ we obtain the optimal weights $\vec{\lambda}^*_{t}\coloneqq\left\{\lambda^*_{2t},\ldots,\lambda_{(J+1)t}\right\}\in\Delta^{J-1}$ by finding the $2$-Wasserstein barycenter of the quantile functions $F_{Y_{jt}}^{-1}$ of the control units which is as close as possible (in the squared $2$-Wasserstein distance) to the quantile function $F_{Y_{1t}}^{-1}$ of the target distribution. Just as in the classical setting, $\Delta^{J-1}$ denotes the $J-1$-unit simplex, which is the set of all $J$-dimensional vectors whose elements are non-negative and sum to unity. But we can also allow for extrapolation beyond the unit simplex as in the classical method by allowing the weights to be negative as long as they sum to unity. This immediately yields a regression approach on the Wasserstein space analogous to the Euclidean setting \citep*{abadie2015comparative}. 

Formally, we compute 
\[\vec{\lambda}^*_t=\argmin_{\vec{\lambda}\in \Delta^{J-1}}\int_0^1 \left\lvert \sum_{j=2}^{J+1}\lambda_j F_{Y_{jt}}^{-1}(q) - F_{Y_{1t}}^{-1}(q)\right\rvert^2dq,\] which we can write as
\begin{equation}\label{weightsuniv}
\vec{\lambda}^*_t=\argmin_{\vec{\lambda}\in\Delta^{J-1}} \left\|\vec{\lambda}^T\mathbb{F}^{-1}_t- F^{-1}_{Y_{1t}}\right\|_{L^2([0,1])}^2,
\end{equation}
where $\mathbb{F}^{-1}_t(q) = \left(F^{-1}_{Y_{2t}}(q),\ldots,F^{-1}_{Y_{(J+1)t}}(q)\right)^T$ is a vector of quantile functions for every $q\in[0,1]$ and $A^T$ denotes the transpose of the matrix $A$.
We could also replace the unit simplex $\Delta^{J-1}$ by the set 
\[\Sigma^{J-1} = \left\{\vec{\lambda}\in\mathbb{R}^J : \vec{\lambda}^T\vec{1} = 1\right\}\] of all weights that sum to unity.\footnote{$\vec{1} = \{1,1,\ldots, 1\}\in\mathbb{R}^J$ denotes the unit vector and $a^Tb$ denotes the inner product in $\mathbb{R}^J$.} This would allow for extrapolation beyond the unit simplex as in the classical method \citep*{abadie2015comparative}, but does not change the method. In this article we focus on the unit simplex as the classical method.

Note that \eqref{weightsuniv} is a simple convex optimization problem for the weights $\lambda^*_t$ which has a unique solution. In particular, in practice one can compute the integral by randomly drawing a large number $M$ of samples $U_m$ from the uniform distribution on the unit interval $U_m\sim U[0,1]$ and solving 
\begin{align*}
\vec{\lambda}^*_t&=\argmin_{\vec{\lambda}\in\Delta^{J-1}} \frac{1}{M}\sum_{m=1}^M\left\lvert\vec{\lambda}^T\mathbb{F}^{-1}_t(U_m)- F^{-1}_{Y_{1t}}(U_m)\right\rvert^2\\
&= \argmin_{\vec{\lambda}\in\Delta^{J-1}} \frac{1}{M}\sum_{m=1}^M\left\lvert\sum_{j=2}^{J+1} \lambda_{jt} F^{-1}_{Y_{jt}}(U_m)- F^{-1}_{Y_{1t}}(U_m)\right\rvert^2
\end{align*} But notice that by the classical inverse transform identity $Y_{jtm} = F^{-1}_{Y_{jt}}(U_m)$, we can write the last expression as a simple linear regression constrained to the unit simplex, i.e.
\begin{equation}\label{regapproach}
\begin{aligned}
\vec{\lambda}^*_t&=\argmin_{\vec{\lambda}\in\Delta^{J-1}} \frac{1}{M}\sum_{m=1}^M\left\lvert\vec{\lambda}^T\mathbb{F}^{-1}_t(U_m)- F^{-1}_{Y_{1t}}(U_m)\right\rvert^2\\
&= \argmin_{\vec{\lambda}\in\Delta^{J-1}} \frac{1}{M}\sum_{m=1}^M\left\lvert\sum_{j=2}^{J+1} \lambda_{jt} Y_{jtm}- Y_{1tm}\right\rvert^2\\
&= \argmin_{\vec{\lambda}\in\Delta^{J-1}} \|\mathbb{Y}_t\vec{\lambda}_t - \vec{Y}_{1t}\|^2_2,
\end{aligned}
\end{equation}
where $\mathbb{Y}_t$ is the $m\times J$-matrix with entry $Y_{jtm}$ at position $(m,j)$, $\vec{Y}_{1t}$ is the vector of elements $Y_{1tm}$ for $m=1,\ldots, M$, and $\|\cdot\|_2$ is the simple Euclidean norm on $\mathbb{R}^m$. Since we approximate the integral by simulations from a uniform distribution we can make the approximation error as small as desired by choosing a sufficiently large number of simulation samples. Since the objective function is convex---which implies that the optimum is well-separated---and continuous in $\vec{\lambda}$, and since $\vec{\lambda}$ is defined on a convex set, the optimal weights obtained from the approximation converge to the optimal weights of the integral expression as $M\to\infty$ \citep*{newey1994large}.

Note that this approach works for any type of distributions, absolutely continuous, discrete, or mixed, as long as they have finite second moments---otherwise the Wasserstein distance can be infinite and the problem becomes trivial. We can then compute the optimal weights $\vec{\lambda}^*$ as a weighted average of the weights $\vec{\lambda}^*_{t}$ over all pre-intervention periods, i.e.
\[\vec{\lambda}^* = \sum_{t\leq T_0} w_{t}\vec{\lambda}^*_t,\qquad\text{for $w_t\geq 0$ and $\sum_{t\leq T_0} w_t = 1$}.\] \citet*{arkhangelsky2019synthetic} provide potential choices of weights $w_t$ that can also be used in our case. In applications, we also found that equal weighs $w_t= \frac{1}{T_0}$ perform well. 

At every time point $t>T_0$ in the post-intervention period, we compute the counterfactual quantile distribution for the treatment unit had it not received the treatment as the optimally weighted barycenter in $2$-Wasserstein space via 
\[F_{Y_{1t,N}}^{-1} = \sum_{j=2}^{J+1}\lambda_{j}^* F_{Y_{jt}}^{-1}.\]

In summary, the whole procedure is as follows.
\begin{definition}[The method of distributional synthetic controls]\label{dscdef}
For any given target distribution $F_{Y_{1t}}$ and control units $\left\{F_{Y_{jt}}\right\}_{j=2,\ldots,J+1}$ observed over $t=1,\ldots, T$ time periods, the method of distributional synthetic controls proceeds as follows:
\begin{enumerate}
\item In the pre-intervention periods, i.e.~for every $t\leq T_0$ perform the following two steps:
\begin{enumerate}
\item Compute the quantile functions $\left\{F_{Y_{jt}}^{-1}\right\}_{j=1,\ldots, J+1}$.
\item Obtain the optimal weights $\vec{\lambda}^*_t$ by solving \eqref{weightsuniv}, potentially via the simple regression approach \eqref{regapproach}.
\item Compute the optimal weights as 
$\vec{\lambda}^* = \sum_{t=1}^{T_0}w_t\vec{\lambda}_t^*$, where $\{w_t\}_{t\leq T_0}$ are some weights in the unit simplex. 
\end{enumerate}
\item In the post-intervention periods, i.e.~for every $T_0<t\leq T$, obtain the counterfactual quantile function $F^{-1}_{Y_{1t},N}$, $t>T_0$, by computing the barycenter of $\{F^{-1}_{Y_{jt}}\}_{j=2,\ldots,J+1}$ for the optimal weight $\vec{\lambda}^*$ obtained in the previous step as:
\[F_{Y_{1t,N}}^{-1}(q) = \sum_{j=2}^{J+1}\lambda_j F_{Y_{jt}}^{-1}(q)\qquad\text{for all $q\in[0,1]$}.\]
\end{enumerate}
\end{definition}
Just as in the classical setting we will in general obtain ``essentially sparse'' weights, i.e.~weights which only put significantly more weight on a few control units. The reason is that we work in a linear space of quantile functions, so that standard linear concepts like convex hulls of quantile functions and linear projections onto the convex hulls make sense. Whenever the target distribution lies outside the convex hull of the donor distribution we obtain sparse weights just as in the classical setting \citep*{abadie2019using}. We showcase this in Figure \ref{univdisc} in section \ref{simulsection}.

The method is easy to implement: run a linear regression to obtain the weights and then take a linear weighted average of the quantile functions. Our method provides the counterfactual quantile function $F_{Y_{1t,N}}^{-1}(q)$ which contains almost all of the information of the distribution function. In particular, if the distribution functions are all continuous and strictly increasing, then one can recover the distribution functions from the quantiles by simply taking the inverse. In more general settings, i.e.~when the distributions are discrete or mixed, the quantiles still provide the information for most important outcomes of interest. For instance, one can compute the expectation $E[Y_{1t,N}]$ as
\[E[Y_{1t,N}] = \int_0^1 F_{Y_{1t,N}}^{-1}(q) dq.\] Based on this one can compute the average treatment effect on the treated unit in time period $t>T_0$ as
\[E[Y_{1t}] - E[Y_{1t,N}] = \int_0^1 F_{Y_{1t}}^{-1}(q) dq - \int_0^1 F_{Y_{1t,N}}^{-1}(q) dq.\] Similarly, one can obtain quantile effects for any quantile by 
\[F_{Y_{1t}}^{-1}(q) - F_{Y_{1t,N}}^{-1}(q).\]
Our method allows for more general statistics too like the counterfactual Lorenz curve \citep*{gastwirth1971general}
\[L_{Y_{1t,N}}(q) = \frac{\int_0^q F^{-1}_{Y_{1t,N}}(t) dt}{\int_0^1 F^{-1}_{Y_{1t,N}}(t) dt},\] Gini-coefficients \citep*{gastwirth1972estimation}, or interquartile ranges, which could be interesting for questions about inequality. Our approach can provide point-wise, i.e.~for fixed $q$, and uniform causal inference on the Lorenz curve using the placebo permutation test Algorithm \ref{CIalgorithm} in section 5. This algorithm is the exact analogue of the classical synthetic controls estimator.

\subsection{Comparison to the classical method and the changes-in-changes estimator}\label{binarysec}
\begin{figure}[h!t]
\centering
\begin{tikzpicture}
\draw node[left] at (-1.5,1.2) {$\boxed{\textbf{DSC}}$};
\draw[->,thick] (-1.5,1.2) to[bend left] (-0.7,1.1);
\draw node[left] at (-0.2,1) {?};
\draw node[left] at (0,-1) {SC};
\draw node[right] at (2.5,-1) {DiD};
\draw node[right] at (2.5,1) {CiC};
\draw[<->] (0,-1) to (2.5,-1);
\draw[<->] (0,1) to (2.5,1);
\draw[->] (-0.5,-0.7) to (-0.5,0.7);
\draw[->] (3,-0.7) to (3,0.7);
\end{tikzpicture}
\caption{Relation between linear and nonlinear synthetic controls and difference-in-differences methods.}
\label{connectionpic}
\end{figure}
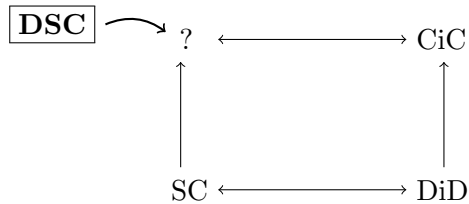

The proposed method completes the schematic in Figure \ref{connectionpic}.
It establishes two existing connections to the existing methods: (i) it is the natural extension of the classical synthetic controls method, and (ii) it is the natural complementary approach to the changes-in-changes estimator in univariate settings. 

\subsubsection{Relation to the classical method}
The proposed method is the correct extension of the classical method in a formal sense: if we apply our distributional method to probability measures supported on just one point, i.e.~Dirac measures of the form
\[\delta_y(A) = \begin{cases} 1& \text{if $y\in A$}\\ 0& \text{otherwise}\end{cases},\] then we obtain the same results as the classical method.\footnote{We provide a proof in the appendix.} This follows from the fact that the setting of one aggregate value can be phrased as a Dirac measure in probability spaces by the standard embedding $x\mapsto \delta_x$ of $\mathbb{R}$ into $\mathcal{P}_2(\mathbb{R})$. This means that the classical method of synthetic controls is a special case of the proposed distributional method: our method reduces to the classical method when we are only given aggregate values and not distributions. 

Note that this does not imply a relation between our method and the classical method applied to moments of a given distribution, such as averages. The classical synthetic controls estimator works with linear averages, which can be computed via
\[E[Y_{1t,N}] = \sum_{j=2}^{J+1}\lambda^*_j E[Y_{jt}].\] The $2$-Wasserstein barycenter is the natural analogues of this for quantile functions:
\[F^{-1}_{Y_{1t,N}}(q)\coloneqq \sum_{j=2}^{J+1}\lambda^*_j F^{-1}_{jt}(q).\] 
This means that obtaining the average for \emph{given weights} $\lambda^*$ in the post-intervention periods our method also implicitly provides an average of the control distributions. In fact, by definition of the expectation as the integral of the quantile function it holds
\[E[Y_{1t,N}]  = \int_0^1\bar{F}^{-1}_{Y_{1t,N}}(q)dq\coloneqq \sum_{j=2}^{J+1}\lambda^*_j \int_0^1F^{-1}_{Y_{jt}}(q)dq = \sum_{j=2}^{J+1}\lambda^*_j E[Y_{jt}].\]
This follows from the fact that the weighted average of quantile functions tries to replicate all moments of the target distribution as closely as possible. 

However, in the pre-intervention periods, the DSC estimator will provide different optimal weights than the classical estimator applied to averages. Intuitively, the reason for this is that the distributional synthetic controls method finds optimal weights that replicate \emph{all} moments of the target distribution as closely as possible. In contrast, applying the classical method to averages of the distributions will find the optimal weights just for the first moment. This can be seen by writing
\begin{align*}
\left\lvert \sum_{j=2}^{J+1}\lambda_j E(Y_{jt}) - E(Y_{1t})\right\rvert &= \left\lvert \sum_{j=2}^{J+1}\lambda_j \int_0^1 F_{Y_{jt}}^{-1}(q) dq- \int_0^1F^{-1}_{Y_{1t}}(q)dq\right\rvert\\
&\leq \int_0^1\left\lvert \sum_{j=2}^{J+1}\lambda_j F_{Y_{jt}}^{-1}(q) - F^{-1}_{Y_{1t}}(q)\right\rvert dq.
\end{align*}
Taking squares on both sides of the expression it follows by the fact that $dq$ is a probability measure on $[0,1]$ in combination with Jensen's inequality that
\[\left( \sum_{j=2}^{J+1}\lambda_j E(Y_{jt}) - E(Y_{1t})\right)^2 \leq \int_0^1\left(\sum_{j=2}^{J+1}\lambda_j F_{Y_{jt}}^{-1}(q) - F^{-1}_{Y_{1t}}(q)\right)^2 dq,\]
where the first expression is the objective function for the classical method when applied to the average of a distribution and the second expression is the objective function for the DSC estimator. This shows, unsurprisingly, that the optimal weights can replicate a single number (the average) at least as well as whole distributions. It also gives an indication that in general the weights will be different for the DSC and the classical method applied to expectations, even in binary settings.

\subsubsection{Relation to the changes-in-changes estimator}
Let us now turn to the connection of our method to the changes-in-changes estimator from \citet*{athey2006identification}. The clearest connection is that the changes-in-changes estimator is also based on the mathematical tools used. In particular, the changes-in-changes estimator relies on the theory of optimal transportation and the Wasserstein space, without explicitly mentioning this connection. Recall the approach of the changes-in-changes estimator: in the case of only one pre- and one post-period, as well as only one control group, they construct the counterfactual distribution $F_{Y_{11,N}}$ as $F_{Y_{10}}(F_{Y_{00}}^{-1}(F_{Y_{01}}))$, where the first index is with respect to the treatment group ($0$ for control and $1$ for treatment) and the second index is with respect to the time period ($0$ is pre and $1$ is post). 

Note in this respect that the monotone rearrangement $F_{Y_{00}}^{-1}(F_{Y_{01}})$ is the optimal transport map between $F_{00}$ and $F_{01}$ with respect to the $2$-Wasserstein distance. In particular, when $F_{Y_{01}}$ is absolutely continuous, one can compute the $2$-Wasserstein distance between $F_{Y_{01}}$ and $F_{Y_{00}}$ as \citep*[Remark 2.19 (iv)]{villani2003topics}
\[W_2(P_{Y_{01}},P_{Y_{00}}) = \left(\int \left\lvert x - F_{Y_{00}}^{-1}(F_{Y_{01}}(x))\right\rvert^2 dF_{Y_{01}}(x)\right)^{1/2}.\] 
Hence the changes-in-changes estimator relies on the exact same mathematical theory of optimal transportation as our proposed method.

Intuitively, \citet*{athey2006identification} estimate the change of the control unit over time and apply this change to the treatment unit.  In contrast, the method of distributional synthetic controls finds the combination of control units that optimally approximates the treatment unit in the pre-intervention periods and then uses this optimal combination of control units to extrapolate the counterfactual distribution in the post period. In a nutshell: the DSC method estimates between units and extrapolates over time, while the changes-in-changes estimator estimates over time and extrapolates between units. 

The proposed method has three main advantages over the changes-in-changes estimator in many settings. First, it provides causal inference for the entire counterfactual distribution or any function of it, such as Lorenz curves, Gini-coefficients, or interquartile ranges. Second, it always provides a unique set of optimal weights $\lambda$ in every case: if the distributions are continuous, discrete, or mixed. In contrast, the changes-in-changes estimator can only provide bounds on average-or quantile effects in the discrete case. 
Third, it does not require the classical rank-invariance assumption \citep*{matzkin2003nonparametric} that the function $h$ is strictly increasing and continuous. 

However, both approaches are complementary: the proposed method does require other relatively strong assumptions on the function $h$ which we point out in the next section. These assumption are at the state-level, not at the individual level, however, which might be more reasonable in practice. Moreover, in contrast to the changes-in-changes estimator, the method works with more than a single control group.

\section{Identification of the counterfactual distribution in a causal model}\label{identsection}
We now introduce assumptions on the causal model $h(t,\cdot)$ for which the obtained counterfactual quantile function $F^{-1}_{Y_{1t, N}}$ via our method coincides with the correct counterfactual quantile function. The restrictions on the causal model become apparent when focusing on the simplest case where we have one pre-intervention period ($t=0$) and one post-intervention period ($t=1$). If we face more than two time-periods, we can adjust the setting by simply assuming that there exists one set $\vec{\lambda}^*$ of optimal weights that hold over all time periods, just as in the classical setting \citep*{abadie2010synthetic}. In this setting, we estimate $\vec{\lambda}^*=\sum_{t=1}^{T_0}w_t\vec{\lambda}^*_t$ for some weights $w_t$, as proposed in \citet*{arkhangelsky2019synthetic}. 

The structure of the synthetic controls method is an extrapolation over time: we obtain the optimal weights $\vec{\lambda}^*_0\equiv \vec{\lambda}^*$ in the pre-intervention period which provide the closest match of the barycenter of the control units to the target unit; the fundamental implicit assumption of synthetic controls approach, be it classical or our proposed generalization, is that these obtained weights $\vec{\lambda}^*$ from the pre-intervention period are still the optimal weights in the post-intervention periods. This assumption has consequences for the functional form of $h(t,U_j)$. 
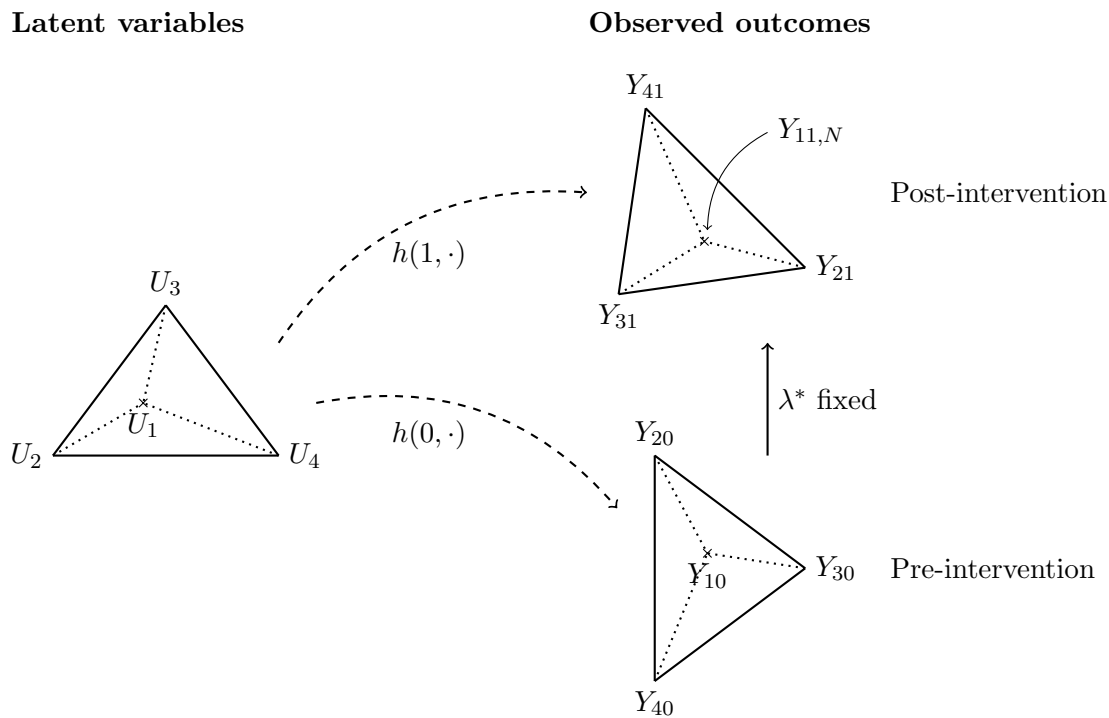
\begin{figure}[h!t]
\centering
\begin{tikzpicture}
\draw node[above] at (1,8) {\textbf{Latent variables}};
\draw node[left] at (0,2.5) {$U_2$};
\draw node[right] at (3,2.5) {$U_4$};
\draw node[above] at (1.5,4.5) {$U_3$};
\draw[-,thick] (0,2.5) -- (3,2.5);
\draw[-,thick] (1.5,4.5) -- (0,2.5);
\draw[-,thick] (1.5,4.5) -- (3,2.5);
\draw node[cross] at (1.2,3.2) {};
\draw node[below] at (1.2,3.2) {$U_1$};
\draw[-,thick,dotted] (0,2.5) to (1.2,3.2);
\draw[-,thick,dotted] (3,2.5) to (1.2,3.2);
\draw[-,thick,dotted] (1.5,4.5) to (1.2,3.2);

\draw node[above] at (9,8) {\textbf{Observed outcomes}};
\draw node[above] at (8,2.5) {$Y_{20}$};
\draw node[below] at (8,-0.5) {$Y_{40}$};
\draw node[right] at (10,1) {$Y_{30}$};
\draw[-,thick] (8,2.5) to (8,-0.5);
\draw[-,thick] (8,-0.5) to (10,1);
\draw[-,thick] (10,1) to (8,2.5);
\draw node[cross] at (8.7,1.2) {};
\draw node[below] at (8.7,1.2) {$Y_{10}$};
\draw[->,thick,dashed] (3.5,3.2) to[bend left] (7.5,1.8);
\draw node[below] at (5, 3.1) {$h(0,\cdot)$};
\draw[thick, dotted] (8,2.5) to (8.7,1.2);
\draw[thick, dotted] (10,1) to (8.7,1.2);
\draw[thick, dotted] (8,-0.5) to (8.7,1.2);
\draw node[right] at (10,5) {$Y_{21}$};
\draw node[above] at (7.88,7.12) {$Y_{41}$}; 
\draw node[below] at (7.52,4.647) {$Y_{31}$}; 
\draw[-,thick] (10,5) to (7.88,7.12);
\draw[-,thick] (7.88,7.12) to (7.52,4.647);
\draw[-,thick] (7.52,4.647) to (10,5);
\draw node[cross] at (8.66,5.35) {};
\draw[-,thick, dotted] (10,5) to (8.66,5.35);
\draw[-,thick, dotted] (7.88,7.12) to (8.66,5.35);
\draw[-,thick, dotted] (7.52,4.647) to (8.66,5.35);
\draw[->,thick,dashed] (3,4) to[bend left] (7.1,6);
\draw node[right] at (9.5,6.8) {$Y_{11,N}$};
\draw[->] (9.5,6.8) to[bend right] (8.7,5.5);
\draw node[below] at (5, 5.5) {$h(1,\cdot)$};
\draw node[right] at (11,1) {Pre-intervention};
\draw node[right] at (11,6) {Post-intervention};
\draw[->,thick] (9.5,2.5) to (9.5,4);
\draw node[right] at (9.5,3.25) {$\lambda^*$ fixed};
\end{tikzpicture}
\caption{Schematic for identification of the counterfactual distribution in model \eqref{modeleq} using a synthetic controls approach with two time periods. The optimal weights $\vec{\lambda}_0^*\equiv\{\lambda^*_{20},\lambda_{30}^*,\lambda_{40}^*\}$ obtained in the pre-intervention period $t=0$ are assumed to still be optimal in the post-intervention period to generate the counterfactual $Y_{11,N}$. A natural sufficient condition for this is that the relative distances between the target $Y_{10}$ and the controls $\{Y_{20},Y_{30},Y_{40}\}$ need to be fixed between $t=0$ and $t=1$. This is achieved by the requirement that the function $h(t,\cdot)$ is a scaled isometry, i.e.~it preserves distances $d(U_1,U_j) = \tau d(h(t,U_1),h(t,U_j)) = \tau d(Y_{1t}, Y_{jt})$ up to a scaling factor $\tau>0$.}
\label{isometrypic}
\end{figure}

Figure \ref{isometrypic} depicts a natural sufficient condition on $h(t,\cdot)$ in model \eqref{modeleq}: the relative distances between the unobservables $U_{1}, U_{2}, U_{3}, U_{4}$ and the observables $Y_{10}, Y_{20}, Y_{30}, Y_{40}$ remain constant over time. A function $h(t,\cdot)$ that preserves distances is an isometry.
\begin{definition}
A map $f:\mathcal{X}\to\mathcal{Y}$ between to metric spaces $(\mathcal{X},d_x)$ and $(\mathcal{Y},d_y)$ is an \emph{isometry} if $d_y(f(x),f(x')) = d_x(x,x')$ for all $x,x'\in\mathcal{X}$. We call $f(x)$ a \emph{scaled isometry} if it satisfies $d_y(f(x),f(x')) = \tau d_x(x,x')$ for some $\tau\in(0,+\infty)$. 
\end{definition}
Isometries are maps which conserve the distance between points, so they are continuous and injective by construction. Under the additional requirement that an isometry is surjective, it becomes invertible.
In one-dimensional Euclidean space, the Mazur-Ulam theorem \citep*{mazur1932transformations} shows that isometries are affine maps, i.e.~compositions of linear functions and shifts. It is here where we can make a connection to the classical method of synthetic controls. In particular, the standard linear factor model for the classical synthetic controls estimator is an affine model which also contains underlying long-run factors. The restriction to isometric/affine models is not surprising as the classical method of synthetic controls requires these restrictions for the extrapolation step going from $t\leq T_0$---where both treatment and control group are observed---to the setting $t>T_0$. 
In our distributional setting, we require that $h$ is a scaled isometry in the $2$-Wasserstein space, which we look at in more detail below. 

Note in this regard that requiring that $h$ be a scaled isometry is only a sufficient assumption. A scaled isometry keeps all distances fixed up to scaling which implies that it preserves barycenters and in particular the weights $\vec{\lambda}_t^*$. However, for our approach to work we only need maps that preserve the weights. These maps can change the overall shape of the convex hull of the control units in general, as long as the target is adjusted so that the optimal weights are preserved. In general, there is not a simple assumption to describe class of functions that preserves the weights $\vec{\lambda}_t^*$, except for isometries. With this assumption we obtain the following result.
\begin{theorem}\label{identthm}
Let the underlying data-generating process be defined by the repeated cross-sections model \eqref{modeleq} where $h(t,\cdot)$ is a scaled isometry on the $2$-Wasserstein space for all $t$. Then the quantile function $F^{-1}_{Y_{1t,N}}$ obtained via method of distributional synthetic controls coincides with the quantile function of the treatment group had it not received the treatment. The analogous result holds for the panel-data model \eqref{modeleq1} if we assume that the maps $U_{jt}\mapsto U_{jt'}$ for $0\leq t,t'\leq T$ and all $j=1,\ldots, J+1$ are such that they preserve the relative weights $\vec{\lambda}^*$ between $P_{U_{1t}}$ and each $P_{U_{jt}}$, $j=2,\ldots, J+1$. 
\end{theorem}
The additional requirement in the panel-data case is that the map that changes the unobservables $U_{jt}$ over time is such that the optimal $\vec{\lambda}^*$ stays the same in the post-treatment periods. This map could also be an isometry, but it need not be.

Since isometries are a natural assumption for $h$, we now review some of their properties. In particular, we introduce the properties of isometries in the $2$-Wasserstein space, i.e.~the space of all distribution functions with finite second moments equipped with the $2$-Wasserstein distance. 
As mentioned, in Euclidean space, scaled isometries are affine functions. It would therefore be a parametric assumption on $h$ if we worked in finite-dimensional Euclidean space. We require $h(t,\cdot)$ to be scaled isometries in the $2$-Wasserstein space, however. 

What do (scaled) isometries look like in the $2$-Wasserstein space? This has been answered in \citet*{kloeckner2010geometric}. Intuitively, isometries in the $2$-Wasserstein space are surprisingly general in the Wasserstein space on the real line, especially compared to simple affine functions \citep*[Theorem 1.1]{kloeckner2010geometric}. In particular, while isometries on the Euclidean space preserve shapes as depicted in Figure \ref{isometrypic}, this is not the case on the Wasserstein space. Intuitively, these isometries on the Wasserstein space only preserve the center of mass of the respective probability distributions and can even deform their support \citep*[section 5]{kloeckner2010geometric}. 

So in general, the admissible ``production functions'' $h(t,\cdot)$ can be more general than simple affine functions for the method of distributional synthetic controls to still work. Assuming linearity in the production function is hence overly restrictive in the setting on the real line (interestingly, in higher dimensions, the admissible isometries are much more restrictive and are only slightly more general than affine functions, see \citet*{kloeckner2010geometric}). This is another extension from the classical synthetic controls method where one assumes a linear model. In particular, the focus on isometries shows that our distributional method can be applied even in the setting with just one pre-intervention and one post-intervention period, analogous to the changes-in-changes estimator \citep*{athey2006identification}. The reason for this is that the assumption of $h$ being scaled isometries is strong enough to not require several time periods for identification. 

\section{Implementation when the distributions are estimated from data}\label{empiricalsection}
In most practical settings the distributions $F_{Y_{jt}}$ are not directly known; instead, only $n\in\mathbb{N}$ random draws $Y_{jti}$, $i=1,\ldots,n$, from each distribution are observed. In this case one needs to first estimate $\left\{F_{Y_{jt}}\right\}_{j=1,\ldots,J+1}$ via standard statistical methods, for instance empirical measures or kernel methods. In this section we provide basic statistical analyses for our method in this section. We provide consistency for our general DSC method and also provide the large sample distribution under regularity assumptions on the distributions. This can be used to obtain (uniform) confidence intervals around the counterfactual quantile function. 

\subsection{The empirical estimator}\label{empestimsection}
The empirical analogue to the optimization \eqref{weightsuniv} for obtaining the weights $\vec{\lambda}^*_t$ for all $t\leq T_0$ is simply the plug-in version
\begin{equation}\label{weightsunivemp}
\hat{\vec{\lambda}}^*_{tn}=\argmin_{\vec{\lambda}\in\Delta^{J-1}} \int_0^1\left\lvert\vec{\lambda}^T\hat{\mathbb{F}}^{-1}_{Y_{tn}}(q)- \hat{F}^{-1}_{Y_{1tn}}(q)\right\rvert^2dq,
\end{equation}
where $\hat{\mathbb{F}}^{-1}_{Y_{tn}}(q) = \left(\hat{F}^{-1}_{Y_{2tn}}(q),\ldots,\hat{F}^{-1}_{Y_{(J+1)tn}}(q)\right)^T$ is a vector of empirical quantile functions and $\hat{F}^{-1}_{Y_{1tn}}(q)$ is the empirical quantile function of the target. 

In the second stage, i.e.~for $t>T_0$, one then simply has to compute the linear weighted average of the empirical quantile functions. We denote the data as $i=1,\ldots, m$ to show that the weights $\hat{\vec{\lambda}}_{n}^*$ were computed on different data. This leads to the following estimator of the counterfactual quantile function.
\[\hat{F}^{-1}_{Y_{1tm,N}}(q) = \sum_{j=2}^{J+1}\hat{\vec{\lambda}}^*_{jn} \hat{F}^{-1}_{Y_{jtm}}(q),\]
where \[\hat{\vec{\lambda}}^*_n=(\hat{\lambda}_{2n}^*,\ldots,\hat{\lambda}_{(J+1)n}^*) = \sum_{t\leq T_0} w_t\hat{\vec{\lambda}}^*_{tn}\] are the optimal weights obtained as a weighted average over time. 

Based on this counterfactual distribution one can obtain any function of interest such as averages, quantiles, Lorenz curves, or other ROC curves \citep*{hsieh1996nonparametric}. 

\subsection{Consistency of the optimal weights $\hat{\vec{\lambda}}_{tn}^*$}\label{weightscons}
Consistency of the optimal weights follows under minimal assumptions. In particular, we only need to assume that the second moments of our estimator exist.
\begin{assumption}\label{univass}
All probability measures have finite second moments, i.e.
\[E[Y_{jt}^2]<+\infty\qquad\text{for $j=1,\ldots, J+1$ and all $t$}\] and are independent across units $j=1,\ldots, J+1$ for each time period $t$.
\end{assumption}

\begin{proposition}[Consistency of the optimal weights]\label{consistencycont}
Let Assumption \ref{univass} hold. Then the corresponding estimator $\hat{\vec{\lambda}}^*_{tn}\in\Delta^{J-1}$ of the optimal weights computed via \eqref{weightsunivemp} converges in probability to the true optimal weights $\vec{\lambda}^*_t$ computed via \eqref{weightsuniv} for all $t\leq T_0$ as the number of observations $n$ approaches infinity.
\end{proposition}

\subsection{Goodness-of-fit tests for post-intervention periods}\label{gofsec}
Working with distributions instead of mere aggregate values as in the classical method has the added benefit that we can empirically test if the estimated distributions via our method are significantly different from the target distribution. In particular, we can test if the counterfactual distribution $F^{-1}_{Y_{1t,N}}$ is statistically different in the Wasserstein distance from the observed outcome $F^{-1}_{Y_{1t}}$. Focusing on the Wasserstein distance is natural. It is a quantile-quantile plot test in the $L^2$-norm \citep*{ramdas2017wasserstein} and takes into account all geometric properties of the distributions when comparing them. 

\subsubsection{Uniform large sample result for smooth distributions}
In order to obtain uniform results for the whole quantile process, we make classical but strong regularity assumptions on the empirical quantile processes (see for instance \citet*{csorgo1993weighted} or chapter 18 in \citet*{shorack2009empirical}). 

\begin{assumption}[Regularity for uniform results in the continuous setting]\label{univass2}
For all $t=1,\ldots,T$ and all $j=1,\ldots, J+1$, the distribution functions $F_{Y_{jt}}$ have finite second moments and are absolutely continuous with a differentiable density $f_{Y_{jt}}$ which is bounded away from zero on its support.
\end{assumption}
Assumption \ref{univass2} is not the weakest possible. In fact, the assumption that the density is bounded away on its support can be slightly relaxed, see \citep*[Theorem 2.1]{csorgo1990distributions}; this leads to more technical assumptions, however, which distract from the main results in this section. More generally, \citet*{knight2002limiting} provides large-sample result for quantiles under more general assumptions. In this case the large-sample distribution and the rate of convergence can be different form the Gaussian case. The required assumptions are however rather high-level.

Under this assumption we now provide (i) uniform confidence intervals for the quantile functions, (ii) the asymptotic distribution of a two-sample test of equality between the counterfactual distribution and the observed distribution of the treatment unit, and (iii) the large sample distributions for a test of first- and second order stochastic dominance for the observed distribution and the counterfactual distribution. \\

\noindent\emph{1. Large-sample distribution for confidence intervals}\\
Under the above assumption we obtain the following result.
\begin{proposition}\label{asymptotprop0}
Let Assumptions \ref{univass} and \ref{univass2} hold. Denote the sample size of corresponding to $F_{Y_{jt}}$ by $n_{jt}$ and assume
\[\frac{n_{jt}}{\sum_{j=1}^{J+1} n_{jt}}\to \gamma_{jt}>0\] for all $j$ and $t$. The large sample distribution of the empirical process of the counterfactual distribution for all $t>T_0$ is 
\begin{multline*}
\sqrt{\frac{n_{1t}\cdots n_{(J+1)t}}{\left(\sum_{j=1}^{J+1} n_{jt}\right)^{J}}}\cdot\thickspace \left(\sum_{j=2}^{J+1}\lambda^*_j \hat{F}_{Y_{jtn_{jt}}}^{-1}(q) - \sum_{j=2}^{J+1}\lambda^*_j F_{Y_{jt}}^{-1}(q)\right)
\\\rightsquigarrow \mathbb{B}(q)\sum_{j=2}^{J+1}\lambda^*_j\sqrt{\prod_{-j}\gamma_{jt}}\frac{1}{f_{Y_{jt}}\left(F^{-1}_{Y_{jt}}(q)\right)},
\end{multline*}
where $\mathbb{B}$ is a standard Brownian bridge on $[0,1]$, $\lambda^*_j$ are the optimal weights which are estimated in earlier time periods, and 
\[\prod_{-j}\gamma_{jt} = \gamma_{1t}\cdots\gamma_{(j-1)t}\cdot\gamma_{(j+1)t}\cdots \gamma_{(J+1)t}.\] Weak convergence is denoted by $\rightsquigarrow$. 
\end{proposition}

Four remarks concerning this result are in order. First, in principle this result allows for testing differences in any continuous functional of the quantile functions, such as averages, quantiles, Lorenz curves, and Gini coefficients. In particular, since Lorenz curves are Hadamard differentiable under weak assumptions \citep*{dobler2019bootstrapping}, the rate of convergence will actually stay the same. Second, note that considering $\lambda^*_j$ as fixed quantities in this setting is fine, as these elements were estimated in earlier periods and their estimators are consistent as shown in Proposition \ref{consistencycont}. Third, the assumption on the sample sizes $n_{jt}$ is standard and simply requires that the samples are asymptotically balanced. If all sample sizes are the same then the rate of convergence simplifies to $n$. Fourth, note that the large sample distribution depends on the unknown distribution $F^{-1}_{Y_{1t}}$. This means that one cannot directly apply this result for testing the equality of distributions. Since the large-sample distribution is Gaussian, one can rely on the standard bootstrap \citep*[chapter 8]{wasserman2013all} to obtain the confidence intervals for the quantile functions in this setting.  \\

\noindent\emph{2. Large-sample distribution for test of equality}\\
We provide the large sample distribution for two-sided tests of equality of the observed distribution $F_{Y_{1t}}$ and the counterfactual distribution $F_{Y_{1t,N}}$ at every $t>T_0$
\begin{equation}\label{hypothesiseq}
H_0: \quad F_{Y_{1t,N}}= F_{Y_{1t}}\qquad \text{versus}\qquad
H_1: \quad F_{Y_{1t,N}}\neq F_{Y_{1t}}.
\end{equation}
We test the hypothesis in the Wasserstein space, i.e.~we use the Wasserstein distance to decide if two distributions are equal or not. The asymptotic distribution under the null hypothesis is the following.
\begin{proposition}\label{asymptotprop1}
Let Assumptions \ref{univass} and \ref{univass2} hold. Denote the sample size of corresponding to $F_{Y_{jt}}$ by $n_{jt}$ and assume
\[\frac{n_{jt}}{\sum_{j=1}^{J+1} n_{jt}}\to \gamma_{jt}>0\] for all $j$ and $t$. Under the null hypothesis in \eqref{hypothesiseq} the large sample distribution of the empirical process of the Wasserstein distance between target and counterfactual distribution for all $t>T_0$ is 
\begin{multline*}
\frac{n_{1t}\cdots n_{(J+1)t}}{\left(\sum_{j=1}^{J+1} n_{jt}\right)^{J}}\cdot\thickspace \int_0^1 \left\lvert \sum_{j=2}^{J+1}\lambda^*_j \hat{F}_{Y_{jtn_{jt}}}^{-1}(q) - \hat{F}_{Y_{1tn_{1t}}}^{-1}(q)\right\rvert^2dq 
\\\rightsquigarrow \int_0^1 (\mathbb{B}(q))^2\left[\sum_{j=2}^{J+1}\lambda^*_j\sqrt{\prod_{-j}\gamma_{jt}}\frac{1}{f_{Y_{jt}}\left(F^{-1}_{Y_{jt}}(q)\right)} - \sqrt{\prod_{-1}\gamma_{jt}}\frac{1}{f_{Y_{1t}}\left(F^{-1}_{Y_{1t}}(q)\right)}\right]^2 dq,
\end{multline*}
where $\mathbb{B}$ is a standard Brownian bridge on $[0,1]$, $\lambda^*_j$ are the optimal weights which are estimated in earlier time periods, and 
\[\prod_{-j}\gamma_{jt} = \gamma_{1t}\cdots\gamma_{(j-1)t}\cdot\gamma_{(j+1)t}\cdots \gamma_{(J+1)t}.\] Weak convergence is denoted by $\rightsquigarrow$. 
\end{proposition}
The same remarks as above are valid for this result. In particular, the large-sample distribution also depends on the unknown distribution. It is possible to adjust the result in \citet*[Theorem 1]{ramdas2017wasserstein} which provides an asymptotic distribution that does not depend on the unknown distribution in the limit, but this requires that all $f_{Y_{jt}}\circ F^{-1}_{Y_{1t}}$, $j=1,\ldots, J+1$ contain the support of $f_{Y_{1t}}\circ F^{-1}_{Y_{1t}}$, which can be a strong assumption in our setting with potentially many units. Moreover, in contrast to the asymptotic distribution in Proposition \ref{asymptotprop0} the large-sample distribution in Proposition \ref{asymptotprop1} is non-Gaussian. 

Therefore, the best way to use Proposition \ref{asymptotprop1} is to perform a Monte-Carlo approach to estimate the asymptotic distribution via a plug-in estimator: estimate all unobserved elements like $f_{Y_{jt}}(F_{Y_jt}^{-1})$ and $\gamma_{jt}$ by their empirical analogues $\hat{f}_{Y_{jtn_{jt}}}(\hat{F}_{Y_jtn_{jt}}^{-1})$ and $\frac{n_{jt}}{\sum_{j=1}^{J+1} n_{jt}}$, sample $B$ random paths from the Brownian bridge process and approximate the integral via some sampling method. For each sample $b=1,\ldots, B$, this provides one value for the test statistic $S_b$. Then one can obtain an approximate $p$-value for the null-hypothesis by comparing the original test statistic $S^0$ to all generated samples $\frac{1}{B}\sum_{b=1}^B\mathds{1}(S^b\geq S^0)$. This provides a simple approximation to the asymptotic $p$-value of the null hypothesis.  \\ 

\noindent\emph{3. Large-sample distribution for tests of stochastic dominance}\\
We can use the same idea for testing stochastic dominance instead of equality. A distribution $F$ (weakly) dominates another distribution $G$ in the first order if $F(x)\leq G(x)$ for all $x\in\mathbb{R}$. $F$ dominates $G$ in the second order if \[\int_{-\infty}^x\left[G(t)-F(t)\right]dt\geq0.\] The concept of stochastic dominance can be pushed to higher orders, but we only focus on the first two orders. Following the straightforward (convex) duality relations between the integrated distribution functions and the corresponding integrated quantile functions \citep*[Theorem 3.1]{ogryczak2002dual}, we can phrase both first- and second-order stochastic dominance in terms of quantile functions. A test for first-order dominance of the observed distribution over the counterfactual can be implemented as
\begin{equation}\label{hypothesisineq1}
\begin{aligned}
&H_0: \quad F^{-1}_{Y_{1t}}(q)\geq F^{-1}_{Y_{1t,N}}(q)\quad\text{for all $q\in(0,1]$}\qquad \text{versus}\\
&H_1: \quad F^{-1}_{Y_{1t}}(q)< F^{-1}_{Y_{1t,N}}(q)\quad\text{for at least one $q\in(0,1]$}.
\end{aligned}
\end{equation}
This hypothesis has become a standard in testing first-order stochastic dominance \citep*{barrett2003consistent}: if the alternative is true then one rejects first-order stochastic dominance. Similarly, we can obtain a test for second order stochastic dominance based on quantile functions \citep*[Theorem 3.2]{ogryczak2002dual} as follows.
\begin{equation}\label{hypothesisineq2}
\begin{aligned}
&H_0: \quad \int_0^q F^{-1}_{Y_{1t}}(s)ds\geq \int_0^q F^{-1}_{Y_{1t,N}}(s)ds\quad\text{for all $q\in[0,1]$}\qquad \text{versus}\\
&H_1: \quad \int_0^q F^{-1}_{Y_{1t}}(s)ds< \int_0^q F^{-1}_{Y_{1t,N}}(s)ds\quad\text{for at least one $q\in[0,1]$}.
\end{aligned}
\end{equation}
Since the null hypotheses in both tests \eqref{hypothesisineq1} and \eqref{hypothesisineq2} are uniform in $q$, we use the $L^2$-norm over $q$ as before, i.e.
\[\int_0^1 \max\left\{F^{-1}_{Y_{1t,N}}(q) - F^{-1}_{Y_{1t}}(q),0\right\}dq\] for the test \eqref{hypothesisineq1} and 
\[\int_0^1 \max\left\{\int_0^q \left[F^{-1}_{Y_{1t,N}}(s) - F^{-1}_{Y_{1t}}(s)\right]ds,0\right\}dq\] for the test \eqref{hypothesisineq2}. The use of the maximum function for obtaining the large-sample distribution has been proposed in other approaches for stochastic dominance testing, in particular \citet*{schmid1996testing}. Both test statistics are positive if the respective null-hypothesis is violated on a set of positive measure. 

 This leads to the following result.
\begin{proposition}\label{dominanceprop}
Let Assumptions \ref{univass} and \ref{univass2} hold. Denote the sample size of corresponding to $F_{Y_{jt}}$ by $n_{jt}$ and assume
\[\frac{n_{jt}}{\sum_{j=1}^{J+1} n_{jt}}\to \gamma_{jt}>0\] for all $j$ and $t$. Under the null hypothesis in \eqref{hypothesisineq1} the large sample distribution for all $t>T_0$ is 
\begin{multline*}
\sqrt{\frac{n_{1t}\cdots n_{(J+1)t}}{\left(\sum_{j=1}^{J+1} n_{jt}\right)^J}}\int_0^1\max\left\{\sum_{j=2}^{J+1}\lambda^*_j\hat{F}^{-1}_{Y_{jtn_{jt}}}(q) - \hat{F}^{-1}_{Y_{1tn_{1t}}}(q),0\right\} dq\\
 \rightsquigarrow \int_0^1\max\left\{\sum_{j=2}^{J+1}\lambda^*_j\sqrt{\prod_{-j}\gamma_{jt}}\frac{\mathbb{B}(q)}{f_{Y_{jt}}\left(F^{-1}_{Y_{jt}}(q)\right)}- \sqrt{\prod_{-1}\gamma_{jt}}\frac{\mathbb{B}(q)}{f_{Y_{1t}}\left(F^{-1}_{Y_{1t}}(q)\right)},0\right\}dq.
\end{multline*}
Under the null hypothesis in \eqref{hypothesisineq2} the large sample distribution for all $t>T_0$ is 
\begin{multline*}
\sqrt{\frac{n_{1t}\cdots n_{(J+1)t}}{\left(\sum_{j=1}^{J+1} n_{jt}\right)^J}}\int_0^1\max\left\{\sum_{j=2}^{J+1}\lambda^*_j\int_0^q\hat{F}^{-1}_{Y_{jtn_{jt}}}(s) ds-\int_0^q\hat{F}^{-1}_{Y_{1tn_{1t}}}(s)ds,0\right\}dq\\
\rightsquigarrow \int_0^1\max\left\{\sum_{j=2}^{J+1}\lambda^*_j\sqrt{\prod_{-j}\gamma_{jt}}\int_0^q\frac{\mathbb{B}(s)}{f_{Y_{jt}}\left(F^{-1}_{Y_{jt}}(s)\right)}ds- \sqrt{\prod_{-1}\gamma_{jt}}\int_0^q\frac{\mathbb{B}(s)}{f_{Y_{1t}}\left(F^{-1}_{Y_{1t}}(s)\right)}ds,0\right\}dq.
\end{multline*}
In both cases $\mathbb{B}$ is a standard Brownian bridge on $[0,1]$, $\lambda^*_j$ are the optimal weights which are estimated in earlier time periods, and 
\[\prod_{-j}\gamma_{jt} = \gamma_{1t}\cdots\gamma_{(j-1)t}\cdot\gamma_{(j+1)t}\cdots \gamma_{(J+1)t}.\] Weak convergence is denoted by $\rightsquigarrow$. 
\end{proposition}

The asymptotic distributions in Proposition \ref{dominanceprop} are analogous to the goodness-of-fit test. In particular, they also depend on the unknown distributions and are therefore not directly useful for statistical testing of the hypotheses. However, as above, they imply that subsampling routines can be used to perform these tests under Assumptions \ref{univass} and \ref{univass2}. The requirements for a $\sqrt{n}$ convergence are strong. It is known that the rate of convergence of the expected Wasserstein distance can be quite slow in cases where the distributions are not smooth enough (\citeauthor*{bigot2018upper} \citeyear{bigot2018upper}, \citeauthor*{bobkov2019one} \citeyear{bobkov2019one}). 

In the case where the respective distributions are not absolutely continuous everywhere (for instance when considering financial returns of cross-sectional data of companies at specific time points, where the overall distribution often has atoms at zero), the above results can only be applied at quantiles $q\in(0,1)$ where Assumption \ref{univass2} is satisfied, i.e.~where $f_{Y_{jt}}(F^{-1}_{Y_{jt}}(q))$ exists and is positive for all $j$. \\

\noindent\emph{4. Large sample distribution for equality in the discrete setting}\\
Many settings of interest in applied research involve discrete distributions. We therefore now propose the uniform large-sample result for the two-sample goodness-of-fit test in this case. The result is a slight extension of Theorem 2.6 in \citet*{samworth2004convergence}. Note also that recently \citet*{sommerfeld2018inference} derived the large sample distribution for more general underlying spaces, but they are designed for fixed distributions and do not take into account the fact that the counterfactual distribution is computed itself. 
\begin{proposition}\label{discasymprop}
Let Assumption \ref{univass} hold and suppose that all probability measures $P_{Y_{jt}}$ are discrete, i.e.~have finite supports $\mathcal{Y}_{jt}$. Denote the sample size of corresponding to $F_{Y_{jt}}$ by $n_{jt}$ and assume
\[\frac{n_{jt}}{\sum_{j=1}^{J+1} n_{jt}}\to \gamma_{jt}>0\] for all $j$ and $t$. Write $\mathcal{Y}_t\coloneqq \bigcup_{j=1}^{J+1} \mathcal{Y}_{jt}$ and denote the respective support points as $\{y_{jts}\}_{s=1,\ldots, S},$ where $S$ is the cardinality of $\mathcal{Y}_t$. Denote by $\{q_s\}_{s=1,\ldots, S}$ all the quantiles where any of the quantile functions $F^{-1}_{Y_{jt}}(q)$, $j=1,\ldots, J+1$, jumps. Define $\{\hat{q}_s\}_{s=1,\ldots, S}$ analogously for the empirical quantile functions $\hat{F}^{-1}_{Y_{jtn_{jt}}}(q)$. Under the null hypothesis in \eqref{hypothesisineq1} the large sample distribution for all $t>T_0$ is
\begin{multline*}
\sqrt{\frac{n_{1t}\cdots n_{(J+1)t}}{\left(\sum_{j=1}^{J+1} n_{jt}\right)^J}}\left\|\sum_{j=2}^{J+1}\lambda_j^*\hat{F}^{-1}_{Y_{jtn_{jt}}} - \hat{F}^{-1}_{Y_{1tn_{1t}}}\right\|_{L^2([0,1])}^2\\\rightsquigarrow \sqrt{\prod_{j=1}^{J+1}\gamma_{jt}}\sum_{s=1}^{S-1} \left\lvert \mathbb{B}(q_s)\right\rvert \left(\sum_{j=2}^{J+1}\lambda^*_j\left(y_{jt(s+1)} - y_{jts}\right) - y_{1t(s+1)} + y_{1ts} \right)^2,
\end{multline*}
where $\mathbb{B}(q)$ is a standard Brownian bridge and $y_{jt(s+1)} = y_{jts}$ for distributions $F_{Y_{jt}}$ for which $y_{jt(s+1)}$ is not a support point. 
\end{proposition}
Proposition \ref{discasymprop} shows that the rate of convergence in this setting is slower than in the regular continuous case, as the rate for the squared Wasserstein distance is now $\sqrt{n}$, not $n$. The same idea of the proof can be used for stochastic dominance results under minimal changes. Proposition \ref{discasymprop} shows that the asymptotic distribution in the discrete case is also non-standard. Therefore, one should resort to Monte-Carlo approaches as mentioned before in the continuous setting. This is in line with the findings from \citet*{sommerfeld2018inference} who show in a multivariate setting that the Wasserstein distance is only directionally Hadamard differentiable, which implies that the standard bootstrap does not work.

\subsubsection{Tests for the counterfactual distribution in pre-intervention periods}
As they stand, the results from the previous section can only be applied in post-intervention periods, as it requires that the optimal weights $\lambda^*$ are known. Since those are computed on independent data-sets in pre-intervention periods via a consistent estimator as shown in Proposition \ref{consistencycont}, it is valid to use the estimator $\hat{\lambda}_{n}^*$ computed over the pre-intervention periods. 

Confidence bands under Assumption \ref{univass2} for the estimated quantile function in the pre-intervention periods can be constructed based on a sample-splitting procedure: since all draws are independently sampled, we can split all samples $\{Y_{jt}\}_{j=1,\ldots,J+1}$ into two independent parts for each $t\leq T_0$. We use the first part to compute the optimal weights $\lambda^*_t$ in each period $t$ and compute the overall optimal $\lambda^* = \sum_{t\leq T_0}w_t\lambda^*_t$ as usual, but only over the first part of the samples that we used to obtain $\lambda^*_t$. On the second sample we can then apply any of the goodness-of-fit tests from the previous sections, in particular the confidence band approach.
The downside of this is that the efficiency in both estimating $\lambda^*$ and constructing the confidence bands is reduced due to the sample split, especially compared to the post-intervention periods, where we can use the whole sample in each period $t$. 
Goodness-of-fit- and stochastic dominance test can be applied via the same sample-splitting idea. 

\subsection{Placebo permutation test}\label{conformalsection}
The goodness-of-fit tests from the previous periods are a necessary condition for a causal effect: if we reject the null, then the distributions in the post-treatment are statistically different. It is not a sufficient condition for a true causal effect. In fact, the synthetic controls estimator might not be able to replicate the target well in the pre-intervention periods, so that a significant difference in the post-intervention periods is not an indication for a true causal effect. One way to gather further evidence for a causal effect is a placebo test, analogous to the classical method \citep*{abadie2010synthetic}: apply our method to each control unit as a placebo permutation test. If our model is correct and there is an actual treatment effect only in the treatment group post-intervention, then our method should estimate the effect for the actual treatment unit to be among the most extreme. The pseudo-algorithm Algorithm \ref{CIalgorithm} provides inference for the counterfactual distribution in a setting with potentially several time periods. 
\begin{algorithm*}
\caption{Placebo tests}
\label{CIalgorithm}
\begin{algorithmic}[1]
\Procedure{Permutation inference for causal inference at times $t\geq T_0$}{}
\For{each unit $\iota=1,\ldots, J+1$}
\For{each time period $t = 1,\ldots,T_0$}
\State obtain and record the optimal weights $\lambda^*_{st,\iota}$ using \eqref{weightsuniv}
\EndFor
\State compute the overall optimal weights $\lambda^*_{s,\iota} = \sum_{t\leq1} w_t\lambda^*_{st,\iota}$ 
\For{each time period $t = T_0+1,\ldots, T$}
\State compute $2$-Wasserstein barycenter using the weights $\lambda^*_{s,\iota}$ to obtain $F^{-1}_{Y_{\iota t,N}}$ 
\State record the distances $d_{\iota t}\coloneqq \int_0^1\left\lvert F^{-1}_{Y_{\iota t,N}}(q)-F^{-1}_{Y_{\iota t,}}(q)\right\rvert^2dq$
\EndFor
\EndFor
\For{each time period $t= T_0+1,\ldots, T$}
\State sort $d_{\iota t}$ decreasingly in $\iota$
\State record the rank $r(d_{1t})$, e.g.~$r(d_{1t})=1$ if $d_{1t}$ is largest
\State compute the probability $p_t$ of obtaining a value of $d_{1t}$ as $p_t = \frac{r(d_{1t})}{J+1}$
\EndFor
\EndProcedure
\end{algorithmic}
\end{algorithm*}
It follows the same underlying idea as the classical placebo-permutation test in \citet*{abadie2010synthetic}. $p_t$ provides the probability of observing a difference between the observable $F^{-1}_{Y_{1t}}$ and the estimated counterfactual $F^{-1}_{Y_{1t,N}}$ given all permutations of the treatment and control groups and hence provides a ``significance level'' for the permutation test.

The complication compared to the classical method is that our outcome of interest $F^{-1}_{Y_{1t,N}}$ is a functional quantity. Thus, in contrast to the classical setting we use the distances 
\[d_{\iota t}\coloneqq \int_0^1\left\lvert F^{-1}_{Y_{\iota t,N}}(q)-F^{-1}_{Y_{\iota t}}(q)\right\rvert^2dq\] to rank the difference between the observed $F_{Y_{\iota t}}$ and the computed $F_{Y_{\iota t,N}}$. These distances are always non-negative. In the classical univariate setting one can obtain a direction of the effect. We also rely on the $2$-Wasserstein distance to measure the distance between the counterfactual and the observed distribution. 
Overall all of these tests can just give an indication of a causal effect, but are not a sufficient condition. It is therefore important to use the method based on a valid model and reasoning to argue that a causal effect exists.

\section{Illustration}\label{simulsection}
This section provides simulations for the core of our proposed method: finding the optimal weights $\lambda_{jt}^*$ for replicating the target distribution at a fixed time period and computing the synthetic control group as the barycenter. In particular, we want to show how well the ``constrained quantile-on-quantile regression'' manages to replicate target distributions. We provide evidence for simulated and real data.

\subsection{Simulated data}
\begin{figure}[h!]
\centering
\textbf{$4$ control distributions}\hspace{3cm} \textbf{$30$ control distributions}\\
\includegraphics[width=7.5cm,height=6cm]{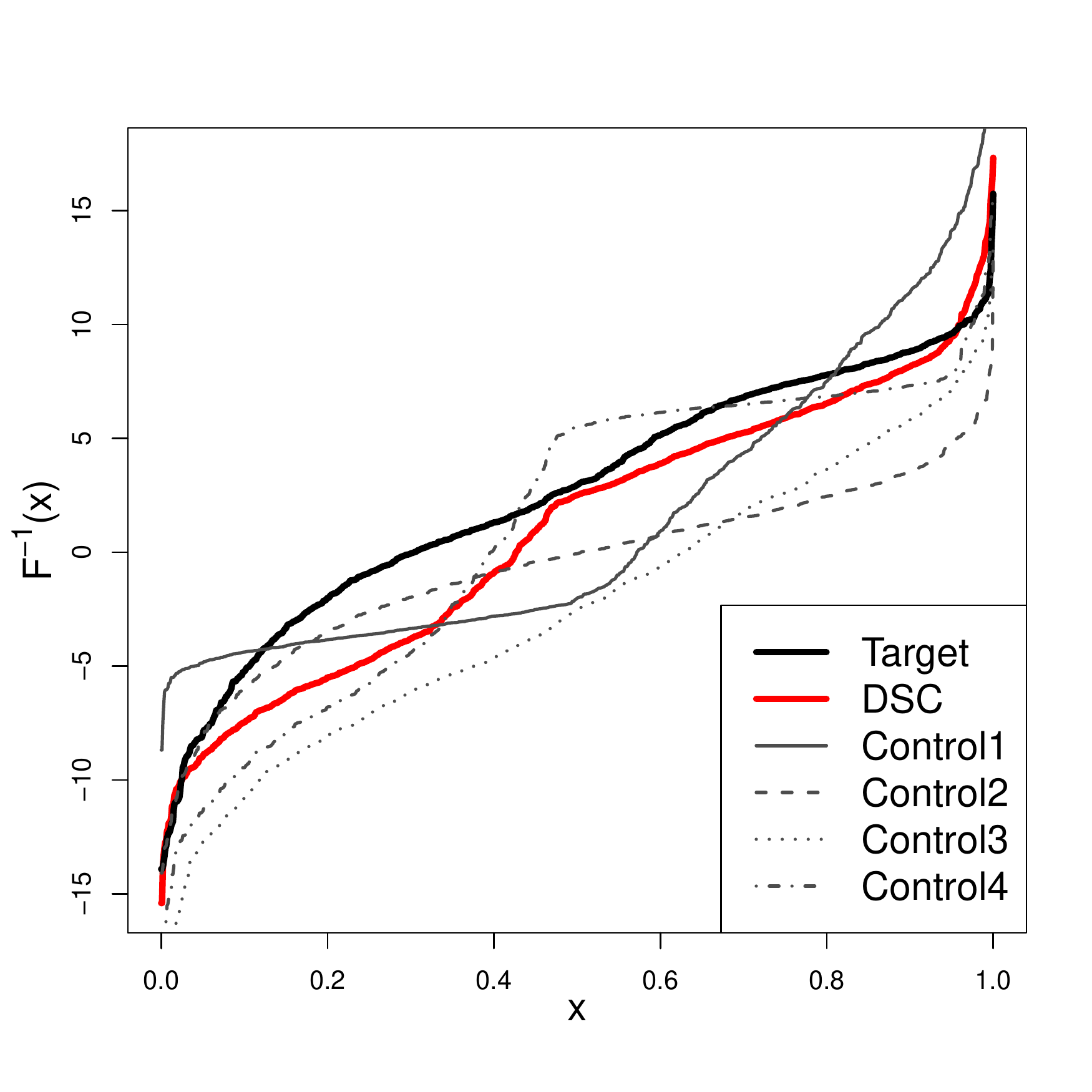}
\includegraphics[width=7.5cm,height=6cm]{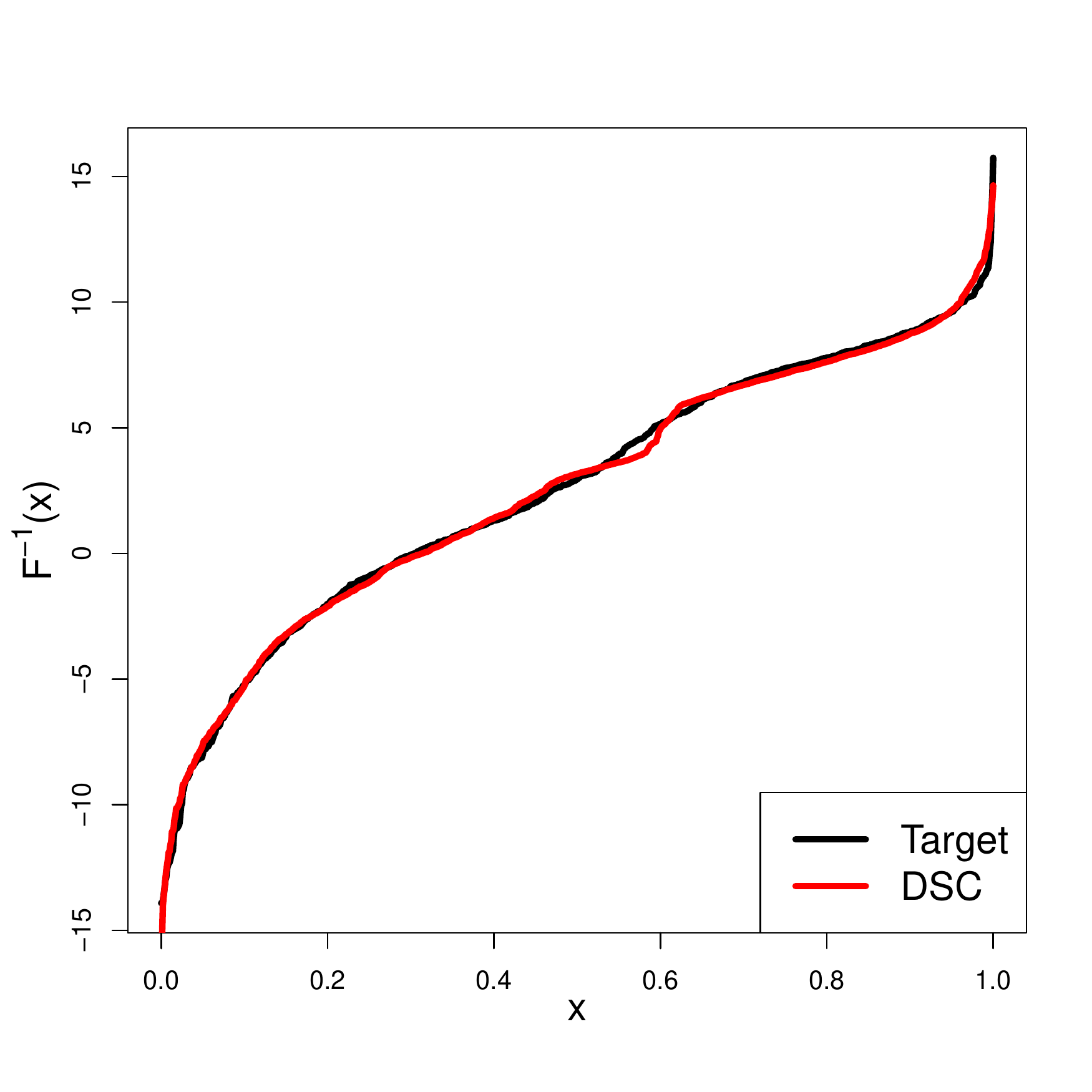}\\
\textbf{$500$ control distributions}\\
\includegraphics[width=7.5cm,height=6cm]{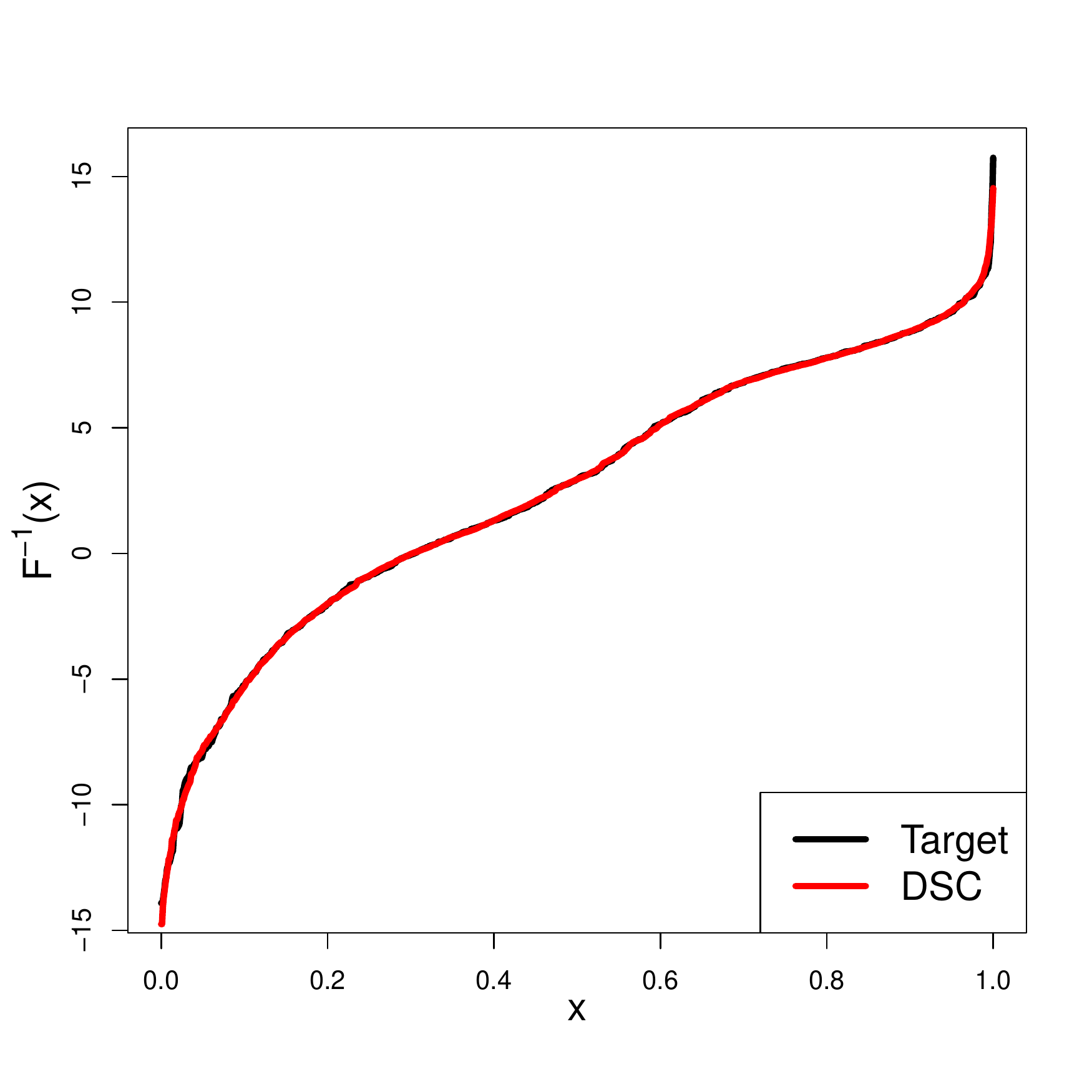}
\caption{Illustration of DSC estimator (thick red) with controls (dashed) randomly generated mixtures of 3 Gaussians and target (thick black) a randomly generated mixture of $4$ Gaussians. $1000$ draws for each distribution.}\label{univmn}
\end{figure}

We first provide simulation evidence in a setting where all distributions are absolutely continuous. We do this by defining the control distributions to be mixtures of $3$ Gaussians, with uniformly randomly generated means between $-10$ and $10$ and variances between $0.5$ and $6$. To maximize the chance that a target that cannot be perfectly replicated by the control units, we define the target to be a mixture of $4$ Gaussians with uniformly randomly generated means between between $-10$ and $10$ and variances between $0.5$ and $6$. Figure \ref{univmn} shows how close the method of distributional synthetic controls can replicate the target for varying numbers of control units in this example. 
It is interesting to note that the method manages to replicate the target rather well with already $30$ control units, and replicates it perfectly with $500$. In fact, much fewer control units are necessary for a perfect replication in this case, but the point is to demonstrate that the method works efficiently with many data points and many control units. In particular, the vector of weights $\vec{\lambda}^*$ is ``essentially sparse'': roughly $5\%$ of the $500$ control units have a weight of greater than $10^{-4}$. It is not formally sparse, however, i.e.~no weight is set to exactly zero. This should not be surprising, as one needs essentially infinite degrees of freedom to replicate an infinite dimensional object like the target, so that even control units with very small weights provide a better fit. However, this essential sparsity shows that one can compress the information of the approach dramatically, which is useful in applied settings. 

\begin{figure}[h!]
\centering
\textbf{$3$ control distributions}\hspace{3cm} \textbf{$4$ control distributions}\\
\includegraphics[width=7.5cm,height=6cm]{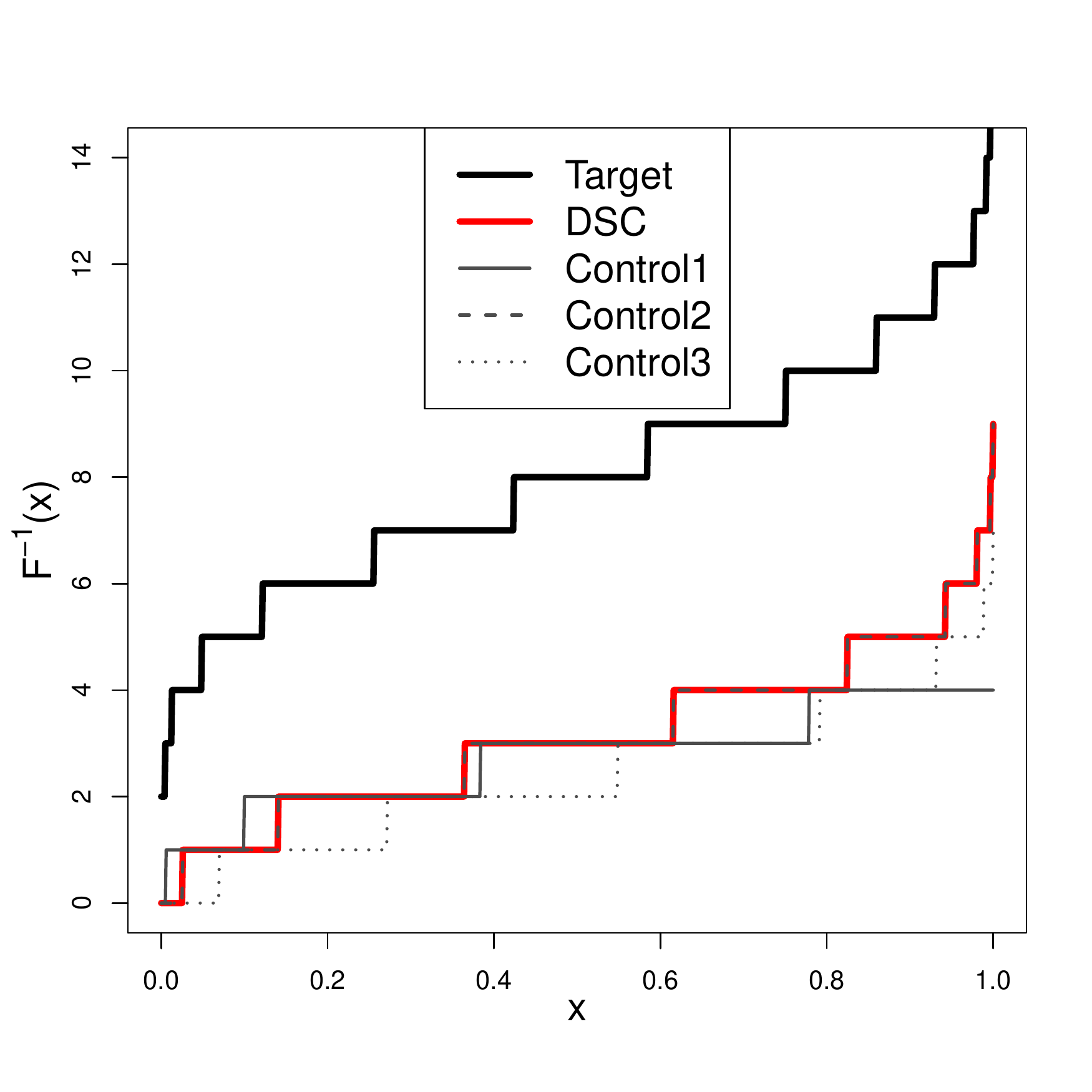}
\includegraphics[width=7.5cm,height=6cm]{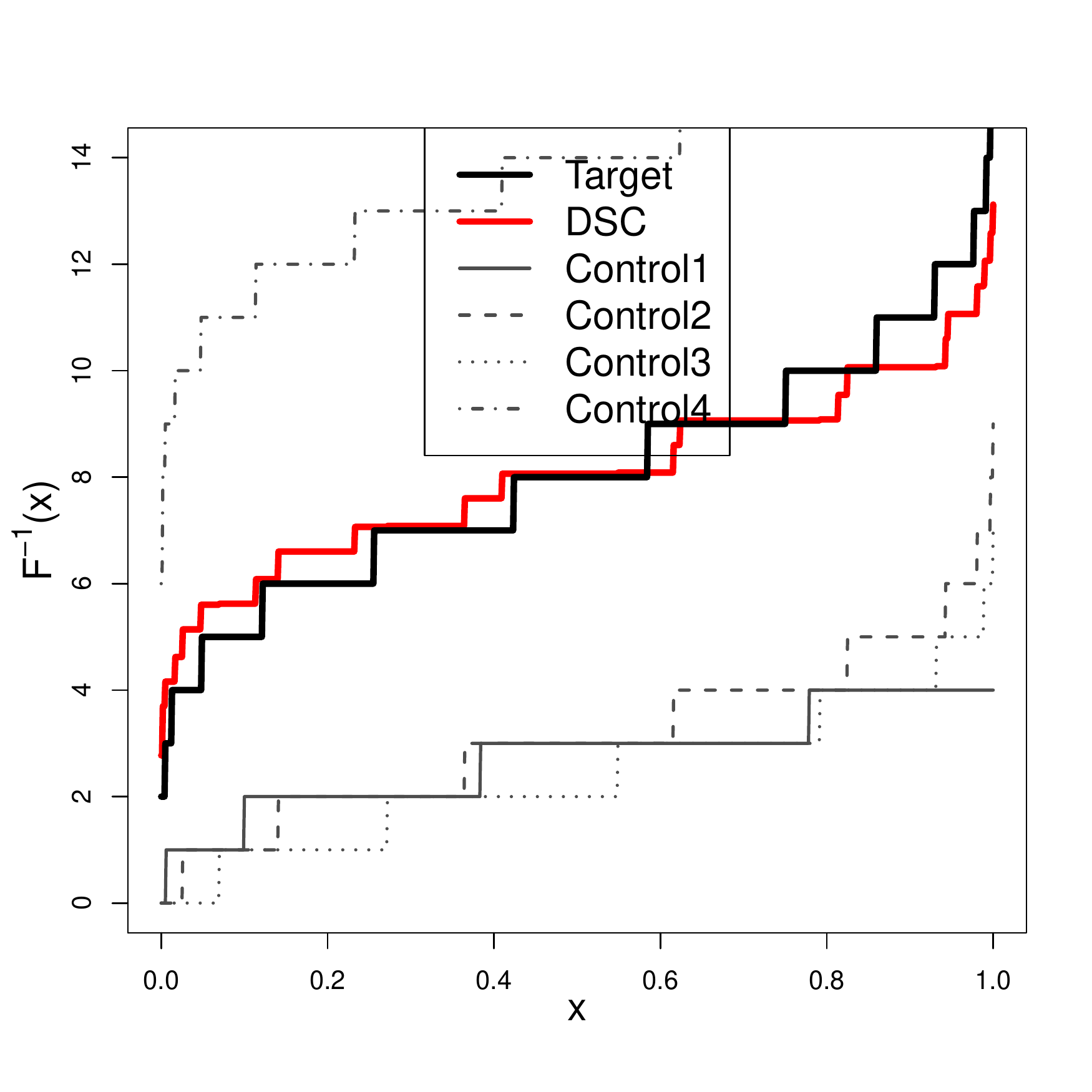}\\
\textbf{$1000$ control distributions}\\
\includegraphics[width=7.5cm,height=6cm]{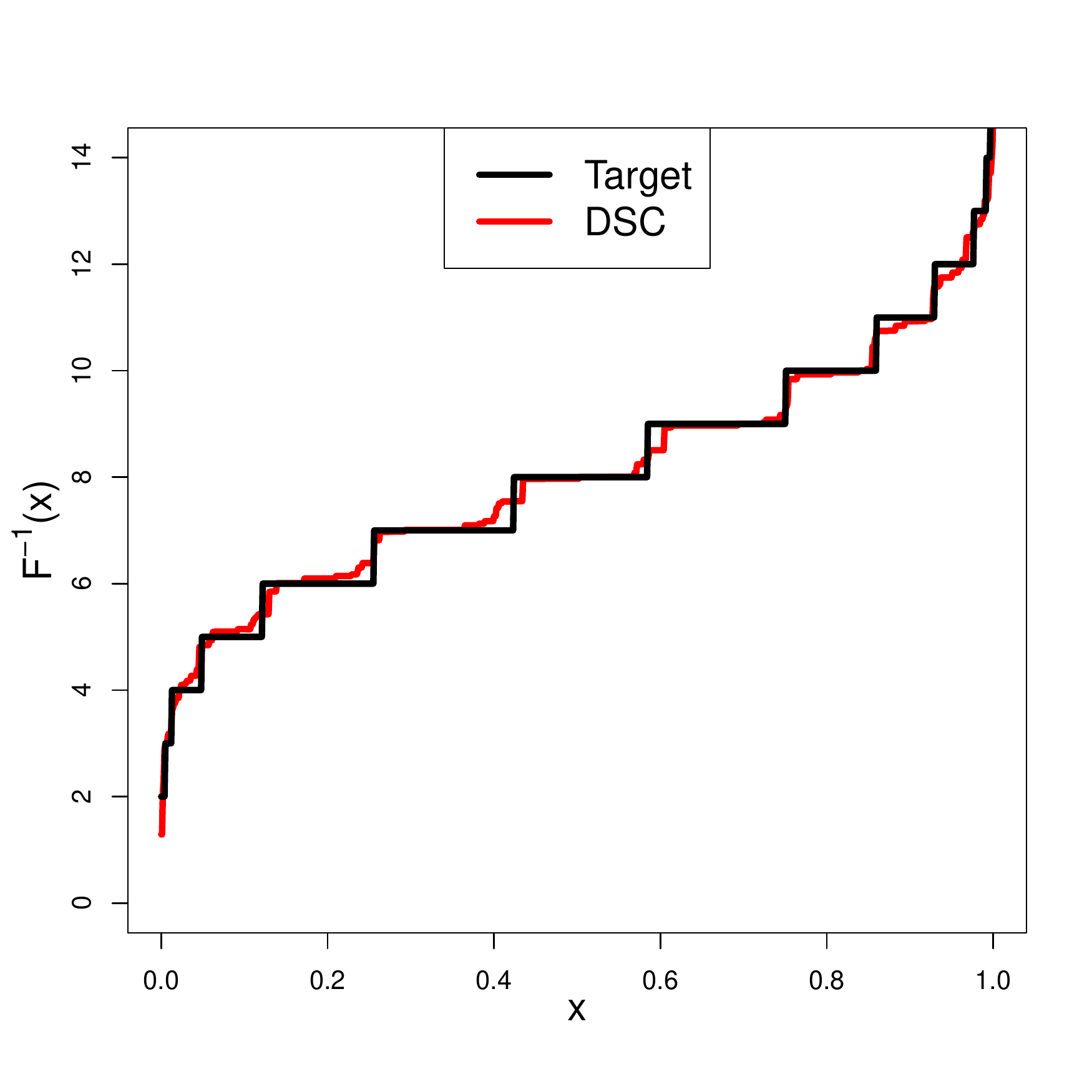}
\caption{Illustration of DSC estimator (thick red) and target (thick black). $1000$ draws for each distribution.}\label{univdisc}
\end{figure}

Even more striking is Figure \ref{univdisc}. It depicts the same setting, but for binomial distributions instead of mixtures of Gaussian. 
Here, the control distributions are very different from the target. However, the ways in which the method is able to replicate the target differ dramatically in each case. In the top-left image the method is quite far from the target while in the top-right image it is able to replicate the target rather well. The reason for this is mathematical. On the right the additional control distribution is placed in such a way that the target lies close to the convex hull of the control distributions. We can define the convex hull as
\[co_W = \left\{F^{-1}: \thickspace F^{-1} = \sum_{j=2}^{J+1}\lambda_j F^{-1}_{Y_{jt}},\quad \thickspace\lambda=(\lambda_2,\ldots, \lambda_{J+1})\in\Delta^{J-1}\right\}.\] It is therefore possible to replicate the target well by the given control distributions even if those distributions are not similar to the target. In the left image the target lies far outside the convex hull of the distributions and it is not possible to accurate replicate it. In the setting with $1000$ control units, the replication is even better, but not perfect, especially compared to the continuous setting in Figure \ref{univmn}. This reflects the fact that the barycenter has a slower rate of convergence in the discrete case (Proposition \ref{discasymprop}) compared to the continuous case (Proposition \ref{asymptotprop1}).

This replication behavior is exactly analogous to the classical method of synthetic controls and relies on the fact that we require all weights to lie in the unit simplex. This means that we only allow for interpolation in our setting, i.e.~we only use the information provided by the control units. One can straightforwardly extend our method to the case where the weights are allowed to be negative as long as they sum to unity. This relaxation then corresponds directly to a linear regression on quantile functions and is perfectly analogous to the regression approach in the classical setting \citep*{abadie2015comparative}. The interpretation of such an extrapolation is an open question and is not as simple to interpret as the interpolation approach. 

\subsection{Real-world data}
\begin{figure}[h!]
\centering
\textbf{Minimum wage data with target MI}\\
\includegraphics[width=7.5cm,height=6cm]{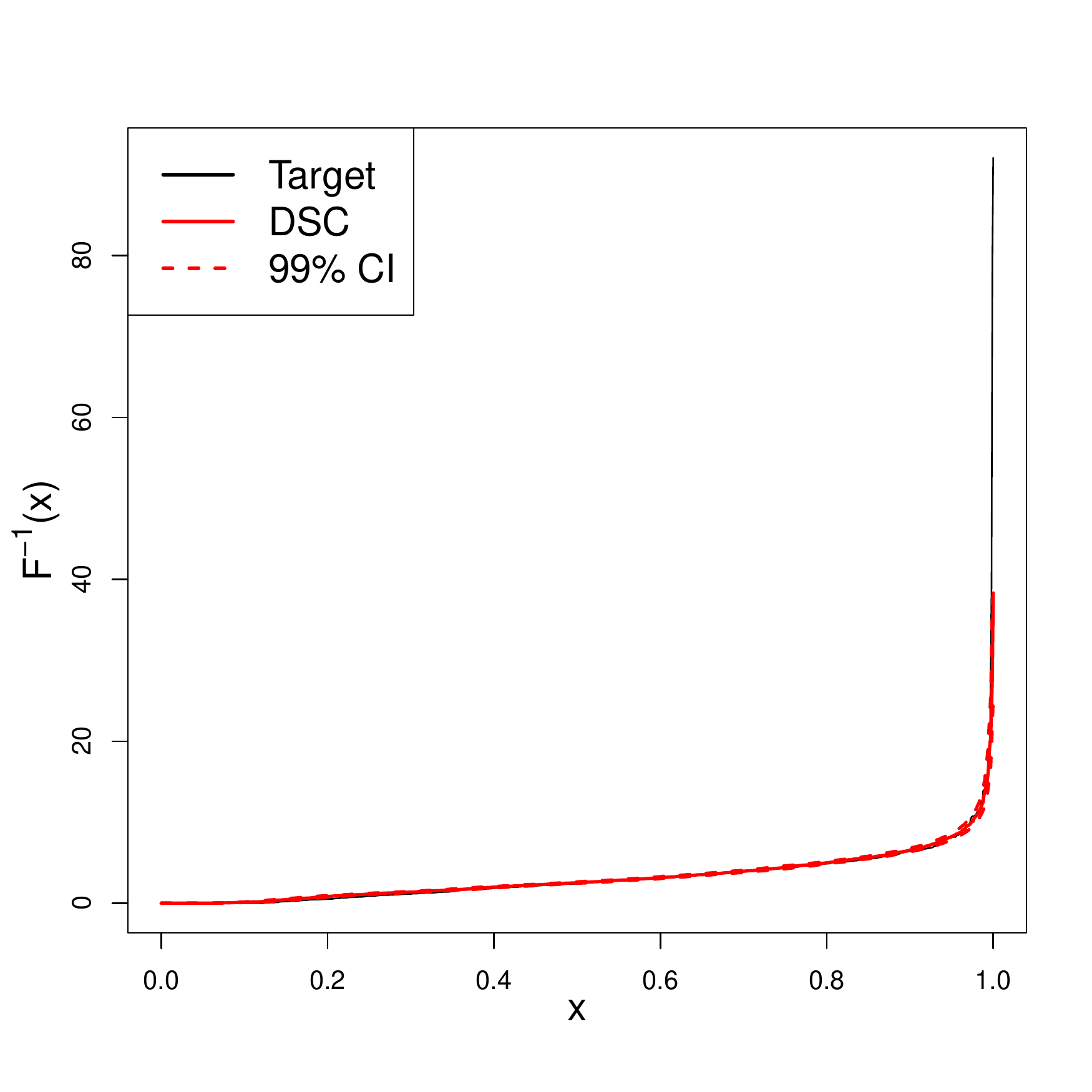}
\caption{Illustration of DSC estimator (thick red) with target Michigan (thick black). Confidence intervals estimated by $50-50$ sample splitting. Optimal weights computed on first half of data, barycenter and confidence intervals estimated on second.}\label{univdata}
\end{figure}

In this section we provide a real-world example for our method. We use a subset of the data provided in \citet*{dube2019minimum} on minimum wages in the United States. The data consist of all $50$ states and the District of Columbia. The outcome of interest is the distribution of equalized family income from wages and salary, defined as multiples of the federal poverty threshold as in \citet*{dube2019minimum}. We pick the final year, $2013$, in the data for illustration. We designate the state of Michigan as the ``treated'' unit we want to replicate and all other $49$ states and DC as control units. Figure \ref{univdata} captures this.

The control units manage to replicate the target perfectly, except for the extreme value for the largest quantile. Figure \ref{univdata} also provides the $99\%$ confidence intervals, but they are so tight that they are hard to distinguish from the estimated line. We estimated this example like in a ``pre-intervention'' period with a $50-50$ data split. On the first sample we estimate the optimal weights $\hat{\vec{\lambda}}$. On the second sample we use those weights to estimate the barycenter and the confidence intervals via a standard bootstrap procedure following the large-sample distribution result from Proposition \ref{asymptotprop0}. Here we also observe the ``essential sparsity phenomenon'': $7$ states received weights larger than $0.0001$. $0.286$ for New Mexico, $0.179$ for Illinois, $0.167$ for Alaska, $0.164$ for Ohio, $0.111$ for Tennessee, $0.084$ for Maryland, and $0.009$ for Massachusetts. 

\section{Conclusion}\label{conclusion}
This article introduced the method of distributional synthetic controls. It is a natural generalization of the classical and widely-used method of synthetic controls (\citeauthor*{abadie2003economic} \citeyear{abadie2003economic} and \citeauthor*{abadie2010synthetic} \citeyear{abadie2010synthetic}) which allows the researcher to replicate whole quantile functions and functional data. The key criterion is to obtain geometrically faithful estimators, i.e.~estimators that approximate the global geometric properties of the target distribution, such as the support. 

We achieve this by relying on the concept of barycenters in Wasserstein spaces \citep*{agueh2011barycenters}. The idea is to find weights such that the corresponding barycenter of the control distributions for these weights is as close as possible to the target. Just like the classical synthetic controls estimator, it requires the weights to lie in the unit simplex, which corresponds to interpolation. One can straightforwardly extend the method to the extrapolation case, analogously to \citep*{abadie2015comparative}, but the interpretability is not clear for these settings. In the latter case, our approach provides the most natural definition of a linear regression of outcomes on Wasserstein space, which defines a mathematical concept that could be of interest in itself. In this article we focus on the method as a tool for causal inference and show in simulations that it performs extraordinarily well in practice, without the need for any tuning parameters. 

Our approach allows us to perform causal inference on many outcomes of interest beyond mere averages or quantiles. For instance, we can perform inference on measures of inequality such as Lorenz curves or Gini-coefficients. Working with distributions instead of aggregate measures allows us to perform goodness-of-fit tests of the hypothesis that the counterfactual is equal to the observed distribution of the treatment. This is a necessary condition for a causal effect, but it is not sufficient. We also provide a placebo permutation approach analogous to  the classical synthetic controls method \citep*{abadie2010synthetic}. In the supplementary material we briefly argue that current existing conformal prediction approaches are not applicable in our setting. It could hence be fruitful to construct conformal prediction approaches for whole probability measures. Another interesting extension of our approach is to allow for staggered adoption.

\bibliography{FSC}

\appendix
\section{Proofs of the results from the main text} 
\subsection{Proof of Theorem \ref{identthm}}
\begin{proof}
Without loss of generality let $\tau_h$, the scaling parameter of the scaled isometry $h$, be equal to unity. The reason is that the scaling does not affect the relative distance between the respective measures which is needed to determining the barycenter. The proof is then straightforward with the definition of an isometry. In particular, as $h(t,\cdot)$ is a surjective isometry in the $2$-Wasserstein space on the real line for all $t$ it holds by definition that 
\[W_2(P_{U_{j}},P_{U_{i}}) = W_2(h_t\sharp P_{U_{j}}, h_t\sharp P_{U_{i}}) = W_2(P_{Y_{jt}},  P_{Y_{it}}) \] for $j,i = 1,\ldots,J+1$, where $h_t(\cdot) \equiv h(t,\cdot)$,  and $h_t\sharp P_{U_{j}}$ denotes the pushforward measure of $P_{U_{j}}$ via $h_t$.\footnote{The pushforward measure of $P_{U_j}$ via $h_t$ is defined as $P_{Y_{jt,N}}(A) = P_{U_j}(h^{-1}_t(A))$ for all (Borel-) sets $A$, where $h^{-1}(A)$ denotes the preimage of the function $h$.} This holds for all time periods $t$. Therefore, the diagram depicted in Figure \ref{isostructure} commutes for all time periods $t,t'\leq T$.
\begin{figure}[h!t]
\centering
\begin{tikzpicture}
\draw node[right] at (3,0) {$P_{Y_{jt}}$};
\draw node[left] at (0,1) {$P_{U_j}$};
\draw node[right] at (3,2) {$P_{Y_{jt'}}$};
\draw node[below] at (1.5,0.1) {$h(t,\cdot)$};
\draw node[above] at (1.6,1.8) {$h(t',\cdot)$};
\draw node[right] at (3.4, 1) {$m_{t,t'}(\cdot)$};
\draw[->,thick] (0,0.9) to (3,0);
\draw[->,thick] (0,1.1) to (3,2);
\draw[->,thick] (3.4,0.4) to (3.4,1.7);
\end{tikzpicture}
\caption{Diagram of isometries}
\label{isostructure}
\end{figure}
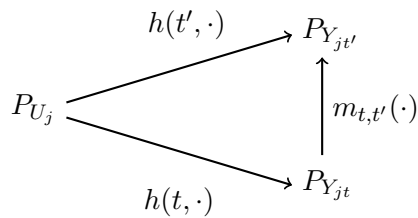
This implies that the map $m_{t,t'}(\cdot)$ is also an isometry for all $t,t'$, as the composition of surjective isometries is a surjective isometry. But since isometries clearly retain barycenters, this implies that using the weights $\lambda_t^*$ obtained in time period $t\leq T_0$ are still optimal in time periods $t>T_0$, so that $P_{Y_{jt,N}}$ obtained by our method provides the correct counterfactual distribution for the model where $h(t,\cdot)$ are isometries. 

The same result holds for model \eqref{modeleq1} when we assume that the weights $\lambda^*$ stay optimal over time for the unobservables $U_{jt}$ even though the relative distances need not. The verbatim argument as above then proves identification for this setting.
\end{proof}

\subsection{Proof of Proposition \ref{consistencycont}}
\begin{proof}
We want to apply the consistency result for M-estimators in Lemma 2.9 in \citet*{newey1994large}. For this we first show that the objective function converges in probability. Then we show that the objective function is ``stochastically H\"older continuous'' as required to apply Lemma 2.9. Compactness and convexity of $\Delta^{J-1}$ and the convexity of the objective function imply the other requirements for Lemma 2.9.\\

\noindent\emph{First step: The objective function converges in probability}\\
We want to show that for every $\varepsilon>0$
\[P\left(\left\lvert\int_0^1 \left\lvert\sum_{j=2}^{J+1}\lambda_j\hat{F}_{Y_{jtn}}^{-1}(q) - \hat{F}_{Y_{1tn}}^{-1}(q)\right\rvert^2dq - \int_0^1 \left\lvert\sum_{j=2}^{J+1}\lambda_jF_{Y_{jt}}^{-1}(q) - F_{Y_{1t}}^{-1}(q)\right\rvert^2dq\right\rvert\geq\varepsilon\right)\] converges to zero. We can write this as
\[P\left(\left\lvert\left\|\sum_{j=2}^{J+1}\lambda_j\hat{F}_{Y_{jtn}}^{-1}-\hat{F}_{Y_{1tn}}^{-1}\right\|_{L^2([0,1])}^2 - \left\|\sum_{j=2}^{J+1}\lambda_jF_{Y_{jt}}^{-1}-F_{Y_{1t}}^{-1}\right\|_{L^2([0,1])}^2 \right\rvert\geq\varepsilon\right).\]
Applying the binomial formula we have
\begin{align*}
&P\left(\left\lvert\left\|\sum_{j=2}^{J+1}\lambda_j\hat{F}_{Y_{jtn}}^{-1}-\hat{F}_{Y_{1tn}}^{-1}\right\|_{L^2([0,1])}^2 - \left\|\sum_{j=2}^{J+1}\lambda_jF_{Y_{jt}}^{-1}-F_{Y_{1t}}^{-1}\right\|_{L^2([0,1])}^2 \right\rvert\geq\varepsilon\right)\\
=&P\left(\left\lvert\sum_{j=2}^{J+1}\lambda_j\left[\left\|\hat{F}_{Y_{jtn}}^{-1}-\hat{F}_{Y_{1tn}}^{-1}\right\|_{L^2([0,1])} - \left\|F_{Y_{jt}}^{-1}-F_{Y_{1t}}^{-1}\right\|_{L^2([0,1])}\right] \right\rvert\right.\cdot\\
&\hspace{2cm}\left. \left\lvert\sum_{j=2}^{J+1}\lambda_j\left[\left\|\hat{F}_{Y_{jtn}}^{-1}-\hat{F}_{Y_{1tn}}^{-1}\right\|_{L^2([0,1])} + \left\|F_{Y_{jt}}^{-1}-F_{Y_{1t}}^{-1}\right\|_{L^2([0,1])}\right] \right\rvert\geq\varepsilon\right)
\end{align*}
Under Assumption \ref{univass} all measures have finite second moments, so that all $2$-Wasserstein distances are finite. This implies that
\begin{align*}
&P\left(\left\lvert\sum_{j=2}^{J+1}\lambda_j\left[\left\|\hat{F}_{Y_{jtn}}^{-1}-\hat{F}_{Y_{1tn}}^{-1}\right\|_{L^2([0,1])} - \left\|F_{Y_{jt}}^{-1}-F_{Y_{1t}}^{-1}\right\|_{L^2([0,1])}\right] \right\rvert\right.\cdot\\
&\hspace{2cm}\left. \left\lvert\sum_{j=2}^{J+1}\lambda_j\left[\left\|\hat{F}_{Y_{jtn}}^{-1}-\hat{F}_{Y_{1tn}}^{-1}\right\|_{L^2([0,1])} + \left\|F_{Y_{jt}}^{-1}-F_{Y_{1t}}^{-1}\right\|_{L^2([0,1])}\right] \right\rvert\geq\varepsilon\right)\\
\leq &P\left(\left\lvert\sum_{j=2}^{J+1}\lambda_j\left[\left\|\hat{F}_{Y_{jtn}}^{-1}-\hat{F}_{Y_{1tn}}^{-1}\right\|_{L^2([0,1])} - \left\|F_{Y_{jt}}^{-1}-F_{Y_{1t}}^{-1}\right\|_{L^2([0,1])}\right]\right\rvert\cdot C \geq\varepsilon\right)
\end{align*}
for some constant $0\leq C<+\infty$. Now using the reverse triangle inequality and the triangle inequality in succession we obtain
\begin{align*}
&P\left(\left\lvert\sum_{j=2}^{J+1}\lambda_j\left[\left\|\hat{F}_{Y_{jtn}}^{-1}-\hat{F}_{Y_{1tn}}^{-1}\right\|_{L^2([0,1])} - \left\|F_{Y_{jt}}^{-1}-F_{Y_{1t}}^{-1}\right\|_{L^2([0,1])}\right]\right\rvert\cdot C \geq\varepsilon\right)\\
\leq &P\left(\sum_{j=2}^{J+1}\lambda_j\left\lvert\left\|\hat{F}_{Y_{jtn}}^{-1}-\hat{F}_{Y_{1tn}}^{-1}\right\|_{L^2([0,1])} - \left\|F_{Y_{jt}}^{-1}-F_{Y_{1t}}^{-1}\right\|_{L^2([0,1])}\right\rvert \geq\frac{\varepsilon}{C}\right)\\
\leq &P\left(\sum_{j=2}^{J+1}\lambda_j\left\|\hat{F}_{Y_{jtn}}^{-1}-F_{Y_{jt}}^{-1}+F_{Y_{1t}}^{-1} - \hat{F}_{Y_{1tn}}^{-1}\right\|_{L^2([0,1])} \geq\frac{\varepsilon}{C}\right)\\
\leq &P\left(\sum_{j=2}^{J+1}\lambda_j\left\|\hat{F}_{Y_{jtn}}^{-1}-F_{Y_{jt}}^{-1}\right\|_{L^2([0,1])}+\left\|F_{Y_{1t}}^{-1} - \hat{F}_{Y_{1tn}}^{-1}\right\|_{L^2([0,1])} \geq\frac{\varepsilon}{C}\right).
\end{align*}

Under Assumption \ref{univass} it follows that \citep*[p.~11]{parzen1980quantile}
\[\int_0^1\left\lvert \hat{F}^{-1}_{Y_{jtn}}(q) - F^{-1}_{Y_{jt}}(q)\right\rvert^2 dq \overset{P}{\to} 0,\] where $\overset{P}{\to}$ denotes convergence in probability. This, in combination with Slutsky's theorem, guarantees that
\begin{align*}
&P\left(\left\lvert\left\|\sum_{j=2}^{J+1}\lambda_j\hat{F}_{Y_{jtn}}^{-1}-\hat{F}_{Y_{1tn}}^{-1}\right\|_{L^2([0,1])}^2 - \left\|\sum_{j=2}^{J+1}\lambda_jF_{Y_{jt}}^{-1}-F_{Y_{1t}}^{-1}\right\|_{L^2([0,1])}^2 \right\rvert\geq\varepsilon\right)\\
\leq &P\left(\sum_{j=2}^{J+1}\lambda_j\left\|\hat{F}_{Y_{jtn}}^{-1}-F_{Y_{jt}}^{-1}\right\|_{L^2([0,1])}+\left\|F_{Y_{1t}}^{-1} - \hat{F}_{Y_{1tn}}^{-1}\right\|_{L^2([0,1])} \geq\frac{\varepsilon}{C}\right) \to 0
\end{align*}
as $n\to\infty$ which is what we wanted to show.\\

\noindent\emph{Step 2: Uniform convergence}\\
We now want to show that the above convergence is uniform over all $\lambda_t\in\Delta^{J-1}$. We use Lemma 2.9 in \citet*{newey1994large}. Since $\Delta^{J-1}$ is convex and compact and the objective function is convex, all that is left to show is a stochastic H\"older condition in the sense that we need to show the existence of a sequence $\hat{B}_n$ of random variables which are bounded in probability and some $\alpha>0$ such that for any $\lambda_t', \lambda_t\in\Delta^{J-1}$ with $\lambda_t' = (\lambda_{2t}',\ldots,\lambda_{2t}')$ and $\lambda_t = (\lambda_{2t},\ldots,\lambda_{2t})$
\[\left\lvert \left\|\sum_{j=2}^{J+1}\lambda'_j\hat{F}_{Y_{jtn}}^{-1}-\hat{F}_{Y_{1tn}}^{-1}\right\|_{L^2([0,1])}^2 - \left\|\sum_{j=2}^{J+1}\lambda_j\hat{F}_{Y_{jtn}}^{-1}-\hat{F}_{Y_{1tn}}^{-1}\right\|_{L^2([0,1])}^2\right\rvert\leq \hat{B}_n\|\lambda'_t-\lambda_t\|_2^\alpha,\]
where $\|\cdot\|_2$ denotes the Euclidean distance on $\mathbb{R}^J$. 
But this follows from a similar argument as before:
\begin{align*}
&\left\lvert \left\|\sum_{j=2}^{J+1}\lambda'_j\hat{F}_{Y_{jtn}}^{-1}-\hat{F}_{Y_{1tn}}^{-1}\right\|_{L^2([0,1])}^2 - \left\|\sum_{j=2}^{J+1}\lambda_j\hat{F}_{Y_{jtn}}^{-1}-\hat{F}_{Y_{1tn}}^{-1}\right\|_{L^2([0,1])}^2\right\rvert\\
=& \left\lvert \left\|\sum_{j=2}^{J+1}\lambda'_j\hat{F}_{Y_{jtn}}^{-1}-\hat{F}_{Y_{1tn}}^{-1}\right\|_{L^2([0,1])} - \left\|\sum_{j=2}^{J+1}\lambda_j\hat{F}_{Y_{jtn}}^{-1}-\hat{F}_{Y_{1tn}}^{-1}\right\|_{L^2([0,1])}\right\rvert\cdot\\
&\hspace{2cm} \left\lvert \left\|\sum_{j=2}^{J+1}\lambda'_j\hat{F}_{Y_{jtn}}^{-1}-\hat{F}_{Y_{1tn}}^{-1}\right\|_{L^2([0,1])} + \left\|\sum_{j=2}^{J+1}\lambda_j\hat{F}_{Y_{jtn}}^{-1}-\hat{F}_{Y_{1tn}}^{-1}\right\|_{L^2([0,1])}\right\rvert\\
\leq&\left\lvert \left\|\sum_{j=2}^{J+1}\lambda'_j\left[\hat{F}_{Y_{jtn}}^{-1}-\hat{F}_{Y_{1tn}}^{-1}\right]\right\|_{L^2([0,1])} - \left\|\sum_{j=2}^{J+1}\lambda_j\left[\hat{F}_{Y_{jtn}}^{-1}-\hat{F}_{Y_{1tn}}^{-1}\right]\right\|_{L^2([0,1])}\right\rvert\cdot \\
&\hspace{2cm} \left\lvert \left\|\sum_{j=2}^{J+1}\lambda'_j\hat{F}_{Y_{jtn}}^{-1}-\hat{F}_{Y_{1tn}}^{-1}\right\|_{L^2([0,1])} + \left\|\sum_{j=2}^{J+1}\lambda_j\hat{F}_{Y_{jtn}}^{-1}-\hat{F}_{Y_{1tn}}^{-1}\right\|_{L^2([0,1])}\right\rvert\\
\leq &\left\|\sum_{j=2}^{J+1}\left[\lambda'_j-\lambda_j\right]\left[\hat{F}_{Y_{jtn}}^{-1}-\hat{F}_{Y_{1tn}}^{-1}\right]\right\|_{L^2([0,1])}\cdot \\
&\hspace{2cm} \left\lvert \left\|\sum_{j=2}^{J+1}\lambda'_j\hat{F}_{Y_{jtn}}^{-1}-\hat{F}_{Y_{1tn}}^{-1}\right\|_{L^2([0,1])} + \left\|\sum_{j=2}^{J+1}\lambda_j\hat{F}_{Y_{jtn}}^{-1}-\hat{F}_{Y_{1tn}}^{-1}\right\|_{L^2([0,1])}\right\rvert\\
\leq & \sum_{j=2}^{J+1}\left\lvert\lambda'_j-\lambda_j\right\rvert\left\|\hat{F}_{Y_{jtn}}^{-1}-\hat{F}_{Y_{1tn}}^{-1}\right\|_{L^2([0,1])}\cdot\\
&\hspace{2cm}\left\lvert \left\|\sum_{j=2}^{J+1}\lambda'_j\hat{F}_{Y_{jtn}}^{-1}-\hat{F}_{Y_{1tn}}^{-1}\right\|_{L^2([0,1])} + \left\|\sum_{j=2}^{J+1}\lambda_j\hat{F}_{Y_{jtn}}^{-1}-\hat{F}_{Y_{1tn}}^{-1}\right\|_{L^2([0,1])}\right\rvert,
\end{align*}
where the equality follows from the binomial formula, the first inequality follows from the fact that the $\lambda_j$ and $\lambda_j'$ sum to unity, the second inequality follows from the reverse triangle inequality, and the third inequality follows from the triangle inequality.
The Cauchy-Bunyakovsky-Schwarz-inequality then implies that
\begin{align*}
& \sum_{j=2}^{J+1}\left\lvert\lambda'_j-\lambda_j\right\rvert\left\|\hat{F}_{Y_{jtn}}^{-1}-\hat{F}_{Y_{1tn}}^{-1}\right\|_{L^2([0,1])}\cdot\\
&\hspace{2cm}\left\lvert \left\|\sum_{j=2}^{J+1}\lambda'_j\hat{F}_{Y_{jtn}}^{-1}-\hat{F}_{Y_{1tn}}^{-1}\right\|_{L^2([0,1])} + \left\|\sum_{j=2}^{J+1}\lambda_j\hat{F}_{Y_{jtn}}^{-1}-\hat{F}_{Y_{1tn}}^{-1}\right\|_{L^2([0,1])}\right\rvert\\
\leq &\|\lambda_t-\lambda'_t\|_2\cdot \hat{B}_n, 
\end{align*}
where
\begin{align*}\hat{B}_n =& \left(\sum_{j=2}^{J+1}\left\|\hat{F}_{Y_{jtn}}^{-1}-\hat{F}_{Y_{1tn}}^{-1}\right\|_{L^2([0,1])}^2\right)^{1/2}\cdot\\ &\hspace{2cm}\left\lvert \left\|\sum_{j=2}^{J+1}\lambda'_j\hat{F}_{Y_{jtn}}^{-1}-\hat{F}_{Y_{1tn}}^{-1}\right\|_{L^2([0,1])} + \left\|\sum_{j=2}^{J+1}\lambda_j\hat{F}_{Y_{jtn}}^{-1}-\hat{F}_{Y_{1tn}}^{-1}\right\|_{L^2([0,1])}\right\rvert
\end{align*}
is the random constant which is bounded in probability under Assumption \ref{univass} as the $2$-Wasserstein distance is finite for distributions with finite second moments. Therefore, the stochastic H\"older condition is satisfied for $\alpha=1$. This, in conjunction with the compactness of $\Delta^{J-1}$, the fact that there exists a unique $\lambda_t^*\in\Delta^{J-1}$ which minimizes the expression in the population due to the convexity of the optimization problem, and that the population analogue is continuous in $\lambda_t$ implies by Theorem 2.1 in \citet*{newey1994large} that 
\[\hat{\lambda}_{tn}^*\overset{P}{\to}\lambda_t^*,\] which is what we wanted to show.
\end{proof}

\subsection{Proof of Proposition \ref{asymptotprop0}}
\begin{proof}
From standard results (e.g.~chapter 18 in \citeauthor*{shorack2009empirical} \citeyear{shorack2009empirical}) it follows under Assumptions \ref{univass} and \ref{univass2} that
\[\sqrt{n_{jt}}\left(\hat{F}^{-1}_{Y_{jtn_{jt}}} - F^{-1}_{Y_{jt}}\right)\rightsquigarrow \frac{\mathbb{B}}{f_{Y_{jt}}\circ F^{-1}_{Y_{jt}}}\] for all $j$ and $t$, where $\mathbb{B}$ is a standard Brownian bridge on $[0,1]$. Since all distributions are independent it follows from the continuous mapping theorem that
\begin{align*}
&\sqrt{\frac{n_{1t}\cdots n_{(J+1)t}}{\left(\sum_{j=1}^{J+1} n_{jt}\right)^J}}\left(\sum_{j=2}^{J+1}\lambda^*_j\hat{F}^{-1}_{Y_{jtn_{jt}}} - \sum_{j=2}^{J+1}\lambda^*_jF^{-1}_{Y_{jtn}}\right)\\
=&\sum_{j=2}^{J+1}\lambda^*_j\sqrt{\frac{\prod_{-j}n_{jt}}{\left(\sum_{j=1}^{J+1} n_{jt}\right)^J}}\left[\sqrt{n_{jt}} (\hat{F}^{-1}_{Y_{jtn_{jt}}}-F^{-1}_{Y_{jt}})\right] \\
 \rightsquigarrow &\sum_{j=2}^{J+1}\lambda^*_j\sqrt{\prod_{-j}\gamma_{jt}}\frac{\mathbb{B}}{f_{Y_{jt}}\circ F^{-1}_{Y_{jt}}}
\end{align*} 

\end{proof}

\subsection{Proof of Proposition \ref{asymptotprop1}}
\begin{proof}
From standard results (e.g.~chapter 18 in \citeauthor*{shorack2009empirical} \citeyear{shorack2009empirical}) it follows under Assumptions \ref{univass} and \ref{univass2} that
\[\sqrt{n_{jt}}\left(\hat{F}^{-1}_{Y_{jtn_{jt}}} - F^{-1}_{Y_{jt}}\right)\rightsquigarrow \frac{\mathbb{B}}{f_{Y_{jt}}\circ F^{-1}_{Y_{jt}}}\] for all $j$ and $t$, where $\mathbb{B}$ is a standard Brownian bridge on $[0,1]$. Since all distributions are independent it follows from the continuous mapping theorem that
\begin{align*}
&\sqrt{\frac{n_{1t}\cdots n_{(J+1)t}}{\left(\sum_{j=1}^{J+1} n_{jt}\right)^J}}\left(\sum_{j=2}^{J+1}\lambda^*_j\hat{F}^{-1}_{Y_{jtn_{jt}}} - \hat{F}^{-1}_{Y_{1tn_{1t}}}\right)\\
=&\sqrt{\frac{n_{1t}\cdots n_{(J+1)t}}{\left(\sum_{j=1}^{J+1} n_{jt}\right)^J}}\left(\sum_{j=2}^{J+1}\lambda^*_j\hat{F}^{-1}_{Y_{jtn_{jt}}} - \sum_{j=2}^{J+1}\lambda^*_jF^{-1}_{Y_{jt}} + F^{-1}_{Y_{1t}} - \hat{F}^{-1}_{Y_{1tn_{1t}}} +\underbrace{\sum_{j=2}^{J+1}\lambda^*_jF^{-1}_{Y_{jt}}-F^{-1}_{Y_{1t}}}_{=0\thickspace\text{under $H_0$}}\right)\\
=&\sum_{j=2}^{J+1}\lambda^*_j\sqrt{\frac{\prod_{-j}n_{jt}}{\left(\sum_{j=1}^{J+1} n_{jt}\right)^J}}\left[\sqrt{n_{jt}} (\hat{F}^{-1}_{Y_{jtn_{jt}}}-F^{-1}_{Y_{jt}})\right] - \sqrt{\frac{\prod_{-1}n_{jt}}{\left(\sum_{j=1}^{J+1} n_{jt}\right)^J}}\left[\sqrt{n_{1t}} (\hat{F}^{-1}_{Y_{1tn_{1t}}}-F^{-1}_{Y_{1t}})\right]\\
 \rightsquigarrow &\sum_{j=2}^{J+1}\lambda^*_j\sqrt{\prod_{-j}\gamma_{jt}}\frac{\mathbb{B}}{f_{Y_{jt}}\circ F^{-1}_{Y_{jt}}}- \sqrt{\prod_{-1}\gamma_{jt}}\frac{\mathbb{B}}{f_{Y_{1t}}\circ F^{-1}_{Y_{1t}}}
\end{align*} 
since the map $L^2([0,1])\ni h\mapsto \int_0^1(h(q))^2dq$ is continuous, another application of the continuous mapping theorem finishes the proof.
\end{proof}

\subsection{Proof of Proposition \ref{dominanceprop}}
\begin{proof}
We first focus on the large-sample distribution for first-order stochastic dominance. \\

\noindent\emph{First setting: large-sample distribution under the null hypothesis of \eqref{hypothesisineq1}.}\\
As in the proof of Proposition \ref{asymptotprop1} it follows under Assumptions \ref{univass} and \ref{univass2} that 
\[\sqrt{n_{jt}}\left(\hat{F}^{-1}_{Y_{jtn_{jt}}} - F^{-1}_{Y_{jt}}\right)\rightsquigarrow \frac{\mathbb{B}}{f_{Y_{jt}}\circ F^{-1}_{Y_{jt}}}\] for all $j$ and $t$, where $\mathbb{B}$ is a standard Brownian bridge on $[0,1]$. Since all samples are iid within and across distributions it follows from the continuous mapping theorem that
\begin{align*}
&\sqrt{\frac{n_{1t}\cdots n_{(J+1)t}}{\left(\sum_{j=1}^{J+1} n_{jt}\right)^J}}\left(\sum_{j=2}^{J+1}\lambda^*_j\hat{F}^{-1}_{Y_{jtn_{jt}}} - \hat{F}^{-1}_{Y_{1tn_{1t}}}\right)\\
=&\sqrt{\frac{n_{1t}\cdots n_{(J+1)t}}{\left(\sum_{j=1}^{J+1} n_{jt}\right)^J}}\left(\sum_{j=2}^{J+1}\lambda^*_j\hat{F}^{-1}_{Y_{jtn_{jt}}} - \sum_{j=2}^{J+1}\lambda^*_jF^{-1}_{Y_{jt}} + F^{-1}_{Y_{1t}} - \hat{F}^{-1}_{Y_{1tn_{1t}}} +\underbrace{\sum_{j=2}^{J+1}\lambda^*_jF^{-1}_{Y_{jt}}-F^{-1}_{Y_{1t}}}_{\leq0\thickspace\text{under $H_0$}}\right)\\
=&\sum_{j=2}^{J+1}\lambda^*_j\sqrt{\frac{\prod_{-j}n_{jt}}{\left(\sum_{j=1}^{J+1} n_{jt}\right)^J}}\left[\sqrt{n_{jt}} (\hat{F}^{-1}_{Y_{jtn_{jt}}}-F^{-1}_{Y_{jt}})\right] - \sqrt{\frac{\prod_{-1}n_{jt}}{\left(\sum_{j=1}^{J+1} n_{jt}\right)^J}}\left[\sqrt{n_{1t}} (\hat{F}^{-1}_{Y_{1tn_{1t}}}-F^{-1}_{Y_{1t}})\right]\\
&\hspace{2cm} +\sqrt{\frac{n_{1t}\cdots n_{(J+1)t}}{\left(\sum_{j=1}^{J+1} n_{jt}\right)^J}}\left(\sum_{j=2}^{J+1}\lambda^*_jF^{-1}_{Y_{jt}}-F^{-1}_{Y_{1t}}\right)\\
 \rightsquigarrow &\begin{cases}\sum_{j=2}^{J+1}\lambda^*_j\sqrt{\prod_{-j}\gamma_{jt}}\frac{\mathbb{B}}{f_{Y_{jt}}\circ F^{-1}_{Y_{jt}}}- \sqrt{\prod_{-1}\gamma_{jt}}\frac{\mathbb{B}}{f_{Y_{1t}}\circ F^{-1}_{Y_{1t}}}& \text{if}\quad \sum_{j=2}^{J+1}\lambda^*_jF^{-1}_{Y_{jt}}=F^{-1}_{Y_{1t}}\\ -\infty &\text{otherwise}.\end{cases}
\end{align*} 
Since $\max\{a,0\} = \frac{a+|a|}{2}$ is continuous, which implies that $L^2([0,1])\ni h\mapsto \int_0^1\max\{h(q),0\}dq$ is continuous, an application of the extended continuous mapping theorem \citep*[Theorem 1.11.1]{wellner2013weak} implies that
\begin{align*}
&\sqrt{\frac{n_{1t}\cdots n_{(J+1)t}}{\left(\sum_{j=1}^{J+1} n_{jt}\right)^J}}\int_0^1\max\left\{\sum_{j=2}^{J+1}\lambda^*_j\hat{F}^{-1}_{Y_{jtn_{jt}}}(q) - \hat{F}^{-1}_{Y_{1tn_{1t}}}(q)\right\} dq\\
 \rightsquigarrow &\int_0^1\max\left\{\sum_{j=2}^{J+1}\lambda^*_j\sqrt{\prod_{-j}\gamma_{jt}}\frac{\mathbb{B}(q)}{f_{Y_{jt}}\left(F^{-1}_{Y_{jt}}(q)\right)}- \sqrt{\prod_{-1}\gamma_{jt}}\frac{\mathbb{B}(q)}{f_{Y_{1t}}\left(F^{-1}_{Y_{1t}}(q)\right)},0\right\}dq.
\end{align*}
\\

\noindent\emph{Second setting: large-sample distribution under the null hypothesis of \eqref{hypothesisineq2}.}\\
As before, under Assumptions \ref{univass} and \ref{univass2} in conjunction with the fact that all distributions are independent and the extended continuous mapping theorem, it holds that
\begin{align*}
&\sqrt{\frac{n_{1t}\cdots n_{(J+1)t}}{\left(\sum_{j=1}^{J+1} n_{jt}\right)^J}}\left(\sum_{j=2}^{J+1}\lambda^*_j\int_0^q\hat{F}^{-1}_{Y_{jtn_{jt}}}(s) ds-\int_0^q\hat{F}^{-1}_{Y_{1tn_{1t}}}(s)ds\right)\\
\rightsquigarrow &\begin{cases}\sum_{j=2}^{J+1}\lambda^*_j\sqrt{\prod_{-j}\gamma_{jt}}\int_0^q\frac{\mathbb{B}(s)}{f_{Y_{jt}}\left(F^{-1}_{Y_{jt}}(s)\right)}ds- \sqrt{\prod_{-1}\gamma_{jt}}\int_0^q\frac{\mathbb{B}(s)}{f_{Y_{1t}}\left(F^{-1}_{Y_{1t}}(s)\right)}ds& \mbox{}\\\hspace{3cm}\text{if}\qquad \sum_{j=2}^{J+1}\lambda^*_j\int_0^qF^{-1}_{Y_{jt}}(s)ds=\int_0^qF^{-1}_{Y_{1t}}(s)ds&\mbox{}\\ -\infty \qquad\qquad\qquad\text{otherwise}&\mbox{}.\end{cases}
\end{align*} 
Applying the extended continuous mapping theorem once more it holds that
\begin{align*}
&\sqrt{\frac{n_{1t}\cdots n_{(J+1)t}}{\left(\sum_{j=1}^{J+1} n_{jt}\right)^J}}\int_0^1\max\left\{\sum_{j=2}^{J+1}\lambda^*_j\int_0^q\hat{F}^{-1}_{Y_{jtn_{jt}}}(s) ds-\int_0^q\hat{F}^{-1}_{Y_{1tn_{1t}}}(s)ds,0\right\}dq\\
\rightsquigarrow &\int_0^1\max\left\{\sum_{j=2}^{J+1}\lambda^*_j\sqrt{\prod_{-j}\gamma_{jt}}\int_0^q\frac{\mathbb{B}(s)}{f_{Y_{jt}}\left(F^{-1}_{Y_{jt}}(s)\right)}ds- \sqrt{\prod_{-1}\gamma_{jt}}\int_0^q\frac{\mathbb{B}(s)}{f_{Y_{1t}}\left(F^{-1}_{Y_{1t}}(s)\right)}ds,0\right\}dq
\end{align*}
\end{proof}

\subsection{Proof of Proposition \ref{discasymprop}}
\begin{proof}
The method for the proof follows the idea of the proof of Theorem 2.6 in \citet*{samworth2004convergence}. Since all probability laws are finite, we denote by $\{q_s\}_{s=1,\ldots, S}$ all the quantiles where any of the quantile functions $F^{-1}_{Y_{jt}}(q)$, $j=1,\ldots, J+1$, jumps. Note that the $q_s$ are ordered, i.e.~$q_1\leq q_2\leq\ldots\leq q_S$. Analogously, we denote by $\{\hat{q}_s\}_{s=1,\ldots,S}$ the quantiles where any of the empirical quantile functions $\hat{F}_{Y_{jtn_{jt}}}^{-1}$, $j=1,\ldots, J+1$, jumps.
$\{q^N_s\}\subset \{q_s\}_{s=1,\ldots, S}$ denotes the set of jump points for the counterfactual quantile function
\[F^{-1}_{Y_{1t,N}}\coloneqq \sum_{j=2}^{J+1}\lambda_j^*F_{Y_{jt}}^{-1},\] where $\vec{\lambda}^*$ are the optimal weights computed in earlier time periods and taken as given. $ \{q_s^I\}\coloneqq \{q_s\}_{s=1,\ldots, S}\setminus \{q^N_s\}$ is the set of quantiles where the quantile function
$F^{-1}_{Y_{1t}}$ jumps. We define $\{\hat{q}^N_s\}$ and $\{\hat{q}^I_s\}$ in the same way. 
Based on this we also denote the supports of the distributions $F_{Y_{jt}}$ as $\mathcal{Y}_{jt}$ and write $\mathcal{Y}_t\coloneqq \bigcup_{j=1}^{J+1} \mathcal{Y}_{jt}$. Note that the relation between the $q_{s}$ and the support points $y_{jts}\in\mathcal{Y}_{jt}$ is that 
\[q_{s_0} = F_{Y_{jt}}(y_{jts_0}),\] where we order the support points $y_{jt1}\leq \ldots\leq y_{jtS}$.  

Define the event 
\[A\coloneqq\left\{\left\lvert \hat{q}_s^N - \hat{q}^I_s\right\rvert\leq\varepsilon\right\}\] for some small enough $\varepsilon>0$, e.g.~$\varepsilon < \frac{1}{3}\min_k\left\lvert q_{s+1}-q_s\right\rvert$.
We now show that its complement converges to an event that has probability zero at the rate of $\sum_{j=1}^{J+1}o\left(n_{jt}^{-1/2}\right)$ under the null hypothesis in \eqref{hypothesiseq}, where $n_{jt}$ are the sample sizes corresponding to the measures $F_{Y_{jt}}$. To see this, note that
\[P(A^c) = P\left(\bigcup_{s=1}^S \left\{\left\lvert \hat{q}_s^N-\hat{q}_s^I\right\rvert>\varepsilon\right\}\right) = P\left(\sup_{y\in\mathbb{R}}\left\lvert \hat{F}_{Y_{1tn_{1t},N}}(y) - \hat{F}_{Y_{1tn_{1t}}}(y)\right\rvert> \varepsilon\right).\]
It also holds that
\begin{align*}
&\sup_{y\in\mathbb{R}}\left\lvert \hat{F}_{Y_{1tn_{1t},N}}(y) - \hat{F}_{Y_{1tn_{1t}}}(y)\right\rvert =  \sup_{y\in\mathbb{R}}\left\lvert \sum_{j=2}^{J+1}\lambda^*_j\hat{F}_{Y_{jtn_{jt}}}(y) - \hat{F}_{Y_{1tn_{1t}}}(y)\right\rvert \\
\leq & \sup_{y\in\mathbb{R}}\left\lvert \sum_{j=2}^{J+1}\lambda^*_j\hat{F}_{Y_{jtn_{jt}}}(y) - \sum_{j=2}^{J+1}\lambda^*_j F_{Y_{jt}}(y)\right\rvert + \sup_{y\in\mathbb{R}}\left\lvert \hat{F}_{Y_{1tn_{1t}}}(y) - F_{Y_{1t}}(y)\right\rvert \\
&\hspace{8cm}+ \underbrace{\sup_{y\in\mathbb{R}}\left\lvert \sum_{j=2}^{J+1}\lambda^*_jF_{Y_{jt}}(y) - F_{Y_{1t}}(y)\right\rvert}_{=0\thickspace\text{under $H_0$}}.
\end{align*}
This implies that the event 
\[A'\coloneqq\left\{\sup_{y\in\mathbb{R}}\left\lvert \sum_{j=2}^{J+1}\lambda^*_j\hat{F}_{Y_{jtn_{jt}}}(y) - \sum_{j=2}^{J+1}\lambda^*_j F_{Y_{jt}}(y)\right\rvert\leq \frac{\varepsilon}{2}\right\}\cap\left\{ \sup_{y\in\mathbb{R}}\left\lvert \hat{F}_{Y_{1tn_{1t}}}(y) - F_{Y_{1t}}(y)\right\rvert \leq\frac{\varepsilon}{2}\right\}\] satisfies $A'\subset A$. Therefore, $A^c\subset (A')^c$, which implies that 
\begin{align*}
&P\left(\sup_{y\in\mathbb{R}}\left\lvert \hat{F}_{Y_{1tn_{1t},N}}(y) - \hat{F}_{Y_{1tn_{1t}}}(y)\right\rvert> \varepsilon\right)\\
\leq &P\left(\sup_{y\in\mathbb{R}}\left\lvert \sum_{j=2}^{J+1}\lambda^*_j\hat{F}_{Y_{jtn_{jt}}}(y) - \sum_{j=2}^{J+1}\lambda^*_j F_{Y_{jt}}(y)\right\rvert > \frac{\varepsilon}{2}\right)+P\left(\sup_{y\in\mathbb{R}}\left\lvert \hat{F}_{Y_{1tn_{1t}}}(y) - F_{Y_{1t}}(y)\right\rvert >\frac{\varepsilon}{2}\right)\\
\leq &P\left(\sum_{j=2}^{J+1}\lambda_j^*\sup_{y\in\mathbb{R}}\left\lvert \hat{F}_{Y_{jtn_{jt}}}(y) -  F_{Y_{jt}}(y)\right\rvert > \frac{\varepsilon}{2}\right)+P\left(\sup_{y\in\mathbb{R}}\left\lvert \hat{F}_{Y_{1tn_{1t}}}(y) - F_{Y_{1t}}(y)\right\rvert >\frac{\varepsilon}{2}\right)
\end{align*} By the fact that $\sum_{j=2}^{J+1}\lambda^*_j=1$ and an analogous argument to the one above we can bound the last line by
\begin{align*}
&P\left(\sum_{j=2}^{J+1}\lambda_j^*\sup_{y\in\mathbb{R}}\left\lvert \hat{F}_{Y_{jtn_{jt}}}(y) -  F_{Y_{jt}}(y)\right\rvert > \frac{\varepsilon}{2}\right)+P\left(\sup_{y\in\mathbb{R}}\left\lvert \hat{F}_{Y_{1tn_{1t}}}(y) - F_{Y_{1t}}(y)\right\rvert >\frac{\varepsilon}{2}\right)\\
\leq &\sum_{j=2}^{J+1}P\left(\sup_{y\in\mathbb{R}}\left\lvert \hat{F}_{Y_{jtn_{jt}}}(y) -  F_{Y_{jt}}(y)\right\rvert > \frac{\varepsilon}{2}\right)+P\left(\sup_{y\in\mathbb{R}}\left\lvert \hat{F}_{Y_{1tn_{1t}}}(y) - F_{Y_{1t}}(y)\right\rvert >\frac{\varepsilon}{2}\right)
\end{align*}

By the Dvoretzky-Kiefer-Wolfowitz-Massart inequality (\citet*{dvoretzky1956asymptotic} and \citet*{massart1990tight}) we can bound this inequality by
\begin{align*}
&\sum_{j=2}^{J+1}P\left(\sup_{y\in\mathbb{R}}\left\lvert \hat{F}_{Y_{jtn_{jt}}}(y) -  F_{Y_{jt}}(y)\right\rvert > \frac{\varepsilon}{2}\right)+P\left(\sup_{y\in\mathbb{R}}\left\lvert \hat{F}_{Y_{1tn_{1t}}}(y) - F_{Y_{1t}}(y)\right\rvert >\frac{\varepsilon}{2}\right)\\
\leq & \sum_{j=1}^{J+1} 2\exp\left(-n_{jt}\frac{\varepsilon^2}{2}\right)
\end{align*}
By the fact that all distributions are discrete it holds for every $j=1,\ldots, J+1$ and every $q\in[0,1]$ that
\[\left\lvert \hat{F}^{-1}_{Y_{jtn_{jt}}}(q) - F^{-1}_{Y_{jt}}(q)\right\rvert\leq \text{diam}(\mathcal{Y}_t),\] where 
\[\text{diam}(\mathcal{Y}_t) = \max_{y,y'\in\mathcal{Y}_t}|y-y'|\] is the diameter of the union of all supports. Squaring both sides and integrating with respect to Lebesgue measure on the unit interval gives
\[\left\|\sum_{j=2}^{J+1}\lambda_j^*\hat{F}^{-1}_{Y_{jtn_{jt}}} - \hat{F}^{-1}_{Y_{1tn_{1t}}}\right\|_{L^2([0,1])}^2\mathds{1}\{A^c\}\leq \text{diam}(\mathcal{Y}_t)\mathds{1}\{A^c\} = o_P\left(\sum_{j=1}^{J+1}n_{jt}^{-1/2}\right).\] where the last part follows from the bound by the Dvoretzky-Kiefer-Wolfowitz-Massart inequality from above. 
This implies that 
\[\left\|\sum_{j=2}^{J+1}\lambda_j^*\hat{F}^{-1}_{Y_{jtn_{jt}}} - \hat{F}^{-1}_{Y_{1tn_{1t}}}\right\|_{L^2([0,1])}^2\mathds{1}\{A^c\}= o_P\left(\left(\sum_{j=1}^{J+1}n_{jt}\right)^{-1/2}\right)\] by the fact that all terms are non-negative and the square-root of a sum is bounded above by the sum the square roots. 

We can therefore focus on the case where $\left\lvert \hat{q}_s^N - \hat{q}^I_s\right\rvert\leq\varepsilon$. In this case, since $\varepsilon < \frac{1}{3}\min_k\left\lvert q_{s+1}-q_s\right\rvert$, we have
\begin{align*}
&\left\|\sum_{j=2}^{J+1}\lambda_j^*\hat{F}^{-1}_{Y_{jtn_{jt}}} - \hat{F}^{-1}_{Y_{1tn_{1t}}}\right\|_{L^2([0,1])}^2\cdot \mathds{1}\{A\} \\
=& \int_0^1\left( \sum_{j=2}^{J+1}\lambda^*_j\hat{F}^{-1}_{Y_{jtn_{jt}}}(q) - \hat{F}^{-1}_{Y_{1tn_{1t}}}(q)\right)^2 dq\\
=& \int_0^1\left(\sum_{j=2}^{J+1}\lambda^*_j\hat{F}^{-1}_{Y_{jtn_{jt}}}(q) - \sum_{j=2}^{J+1}\lambda^*_j F^{-1}_{Y_{jt}}(q) + F^{-1}_{Y_{1t}}(q) - \hat{F}^{-1}_{Y_{1tn_{1t}}}(q) + \underbrace{\sum_{j=2}^{J+1}\lambda^*_j F^{-1}_{Y_{jt}}(q) - F^{-1}_{Y_{1t}}(q)}_{=0\thickspace\text{under $H_0$}}\right)^2 dq\\
= & \int_0^1\left( \sum_{j=1}^{J+1}\tilde{\lambda}^*_j\left(\hat{F}^{-1}_{Y_{jtn_{jt}}}(q) - F^{-1}_{Y_{jt}}(q)\right)\right)^2 dq,
\end{align*}
Where we have defined $\vec{\tilde{\lambda}}^* \equiv (-1, \vec{\lambda}^*)$, i.e.~we extend the weights by $\lambda_1^* = -1$ for notational convenience.

Since all distributions have finite support, the integral over $q$ reduces to a finite sum, that is
\[\int_0^1\left(\sum_{j=1}^{J+1}\tilde{\lambda}^*_j\left(\hat{F}^{-1}_{Y_{jtn_{jt}}}(q) - F^{-1}_{Y_{jt}}(q)\right)\right)^2 dq=\sum_{s=1}^{S-1} \left\lvert \hat{q}_s - q_s\right\rvert \left(\sum_{j=1}^{J+1}\tilde{\lambda}^*_j\left(y_{jt(s+1)} - y_{jts}\right)\right)^2,\]
where the quantiles $\hat{q}_s\geq 0$ are drawn with respect to all data $\sum_{j=1}^{J+1} n_{jt}$ and the points $y_{jts}$ are all support points of $\mathcal{Y}_{t}$. Now by construction these empirical quantiles $\hat{q}_s$ are sorted draws from a multinomial distribution and satisfy the following central limit theorem (see \citet*{samworth2004convergence} and Corollary 1 in \citet*{sommerfeld2018inference})
\[\sqrt{\sum_{j=1}^{J+1} n_{jt}}\left(\hat{q}_s - q_s\right)\rightsquigarrow N(0,\Sigma),\]
where the $rs$-th entry $\Sigma_{rs}$ of the covariance matrix $\Sigma$ is $q_r(1-q_s)$. This is the distribution of a standard Brownian bridge $B(q)$ supported on the points $q_r$ and $q_s$.

Therefore, by the continuous mapping theorem
\begin{align*}
&\sqrt{\frac{n_{1t}\cdots n_{(J+1)t}}{\left(\sum_{j=1}^{J+1} n_{jt}\right)^J}}\left\|\sum_{j=2}^{J+1}\lambda_j^*\hat{F}^{-1}_{Y_{jtn_{jt}}} - \hat{F}^{-1}_{Y_{1tn_{1t}}}\right\|_{L^2([0,1])}^2\\
=&\sqrt{\frac{n_{1t}\cdots n_{(J+1)t}}{\left(\sum_{j=1}^{J+1} n_{jt}\right)^J}}\left\|\sum_{j=2}^{J+1}\lambda_j^*\hat{F}^{-1}_{Y_{jtn_{jt}}} - \hat{F}^{-1}_{Y_{1tn_{1t}}}\right\|_{L^2([0,1])}^2\cdot \mathds{1}\{A\} \\
&\hspace{3cm}+ \sqrt{\frac{n_{1t}\cdots n_{(J+1)t}}{\left(\sum_{j=1}^{J+1} n_{jt}\right)^J}}\left\|\sum_{j=2}^{J+1}\lambda_j^*\hat{F}^{-1}_{Y_{jtn_{jt}}} - \hat{F}^{-1}_{Y_{1tn_{1t}}}\right\|_{L^2([0,1])}^2\cdot \mathds{1}\{A^c\}\\
=&\sqrt{\frac{n_{1t}\cdots n_{(J+1)t}}{\left(\sum_{j=1}^{J+1} n_{jt}\right)^J}}\left\|\sum_{j=2}^{J+1}\lambda_j^*\hat{F}^{-1}_{Y_{jtn_{jt}}} - \hat{F}^{-1}_{Y_{1tn_{1t}}}\right\|_{L^2([0,1])}^2\cdot \mathds{1}\{A\} +o_P(1)\cdot \sqrt{\prod_{j=1}^{J+1}\gamma_{jt}}\\
=&\sqrt{\frac{n_{1t}\cdots n_{(J+1)t}}{\left(\sum_{j=1}^{J+1} n_{jt}\right)^J}} \sum_{s=1}^{S-1} \left\lvert \hat{q}_s - q_s\right\rvert \left(\sum_{j=1}^{J+1}\tilde{\lambda}^*_j\left(y_{jt(s+1)} - y_{jts}\right)\right)^2\\
=& \sqrt{\frac{n_{1t}\cdots n_{(J+1)t}}{\left(\sum_{j=1}^{J+1} n_{jt}\right)^{J+1}}} \sqrt{\sum_{j=1}^{J+1} n_{jt}} \sum_{s=1}^{S-1} \left\lvert \hat{q}_s - q_s\right\rvert \left(\sum_{j=1}^{J+1}\tilde{\lambda}^*_j\left(y_{jt(s+1)} - y_{jts}\right)\right)^2\\
\rightsquigarrow & \sqrt{\prod_{j=1}^{J+1}\gamma_{jt}}\sum_{s=1}^{S-1} \left\lvert \mathbb{B}(q_s)\right\rvert \left(\sum_{j=1}^{J+1}\tilde{\lambda}^*_j\left(y_{jt(s+1)} - y_{jts}\right)\right)^2,
\end{align*}
where $\mathbb{B}(q)$ is a standard Brownian bridge. 
\end{proof}

\section{Additional results}
\subsection{The proposed method reduces to the classical method for Dirac measures}
Suppose we are given aggregate values $y_{jt}$, $j=1,\ldots, J+1$, for all time periods. We can embed this into the distributional setting via $y_{jt}\mapsto \delta_{y_{jt}}$. We now show that the first step of the method reduces to the classical method in this case. 
For this note that the optimal weights in general should be computed via
\[\vec{\lambda}^*_t=\argmin_{\vec{\lambda}\in \Delta^{J-1}}\int_0^1 \left\lvert \sum_{j=2}^{J+1}\lambda_j F_{Y_{jt}}^{-1}(q) - F_{Y_{1t}}^{-1}(q)\right\rvert^2dq.\]
Since the quantile functions $F_{Y_{jt}}$ correspond to Dirac measures centered at some $y_{jt}$ by assumption, they map every $q\in[0,1]$ to their respective point $y_{jt}$. This means that the optimization reduces to 
\[\argmin_{\vec{\lambda}_{t}\in\Delta^{J-1}} \left\lvert\sum_{j=2}^{J+1} \lambda_{jt} y_{jt}- y_{1t}\right\rvert^2\]
which coincides with the classical synthetic controls estimator. The same results holds for the second stage when comparing $\sum_{j=2}^{J+1}\lambda^*_j y_{jt}$ and $\sum_{j=2}^{J+1}\lambda^*_j F^{-1}_{Y_{jt}}$ for Dirac measures.

\subsection{Inadequacy of existing conformal inference approaches}
Recently, conformal prediction approaches (\citeauthor*{gammerman1998learning} \citeyear{gammerman1998learning}, \citeauthor*{shafer2008tutorial} \citeyear{shafer2008tutorial}, \citeauthor*{vovk2005algorithmic} \citeyear{vovk2005algorithmic}) have been introduced for the classical synthetic controls method, such as \citet*{chernozhukov2017exact}, \citet*{chernozhukov2018exact}, and \citet*{cattaneo2019prediction}. These approaches are geared towards finite-dimensional outcomes, while our method relies on infinite-dimensional concepts. Because of this these methods are not applicable in our case, even if we only focus on functionals $f(P_{Y_{1t,N}}, P_{Y_{1t}})$ of the counterfactual distribution. 

To be more precise, consider the arguably most promising approach for our setting in \citet*{chernozhukov2017exact}. The conceptual idea is simple: for a univariate outcome of interest $Y_{1t}$ observed over time periods $t=1,\ldots, T$, one constructs a potential treatment effect sequence $\alpha_t = Y_{1t}-Y_{1t,N}$ with $\alpha_t=0$ for $t\leq T_0$ and tests the null hypothesis that the estimated trajectory coincides with the postulated one. Now---and this is the crucial part for why this approach does not work in our setting---one needs to obtain the optimal weights $\lambda^*_t$ for the method by taking into account all time periods $t=1,\ldots T$. In particular, this means that one needs to include the postulated counterfactuals $P_{Y_{1t,N}}$ when calculating the weights. But since we are in a functional setting it is not clear how to define a potential treatment sequence in general, as a definition like $P_{Y_{1t,N}} \in\{P\in \mathcal{P}_2: d(P_{Y_{1t}},P)=\alpha_t\}$ for some metric $d$ defines a set of potential treatment effect sequences which are very different from each other. Also just assuming that the counterfactual is a shift of the whole distribution is a too strong assumption. 

Note that the weights $\lambda^*_t$ obtained over all time periods $1\leq t\leq T$ are required even when we consider some functional $f(P_{Y_{1t,N}}, P_{Y_{1t}})$, as we always need them to compute the counterfactuals. A similar argument holds for the other conformal prediction approaches which are less suited for our purposes. The extension of classical conformal prediction approaches to functional data as introduced in \citet*{lei2015conformal} is also not applicable in a general time-series setting such as the one for synthetic controls. Furthermore, it is designed for a basis expansion of the functional quantity. This is not useful in our quantile approach where one needs the precise probabilities and not just approximations.

\end{document}